\documentclass[12pt, draftclsnofoot, onecolumn]{IEEEtran}

\IEEEoverridecommandlockouts

\usepackage{graphicx,cite,acronym,amsmath,amssymb,amsfonts,color,subfigure,floatflt,stfloats,bm,textgreek}
\usepackage{epsfig,epstopdf,psfrag}
\usepackage[ruled,vlined]{algorithm2e}
\SetKwInput{KwInput}{Input}                
\SetKwInput{KwOutput}{Output}  

\usepackage[table]{xcolor}
\usepackage{bm}

\usepackage[cmintegrals]{newtxmath}
\usepackage{graphics,hhline,array,cite,bbm}

\usepackage{latexsym}
\usepackage{theorem}
\usepackage{times}
\usepackage{makeidx}

\DeclareGraphicsExtensions{.png,.jpg,.eps}

\acrodef{IIoT}{industrial Internet-of-things}

\acrodef{SRE}{smart radio environment}

\acrodef{EMO}{electromagnetic object}

\acrodef{SVD}{singular value decomposition}

\acrodef{PSWF}{prolate spheroidal wave function}

\acrodef{CR}{channel response}

\acrodef{BS}{base station}

\acrodef{MS}{mobile station}

\acrodef{UE}{user equipment}

\acrodef{MIMO}{multiple-input multiple-output}

\acrodef{RIS}{reconfigurable intelligent surface}

\acrodef{IRS}{intelligent reconfigurable surface}

\acrodef{LIS}{large intelligent surface}

\acrodef{MIS}{medium intelligent surface}

\acrodef{SIS}{small intelligent surface}

\acrodef{DoF}{degrees-of-freedom}

\acrodef{AF}{amplify \& forward}

\acrodef{DF}{detect \& forward}

\acrodef{JF}{just forward}

\acrodef{CSI}{channel state information}

\acrodef{RV}{random variable}

\acrodef{i.i.d.}{independent, identically distributed}

\acrodef{PSD}{power spectral density}

\acrodef{PDF}{probability distribution function}

\acrodef{CDF}{cumulative distribution function}

\acrodef{ch.f.}{characteristic function}

\acrodef{AWGN}{additive white Gaussian noise}

\acrodef{RSSI}{received signal strength indicator}

\acrodef{SNR}{signal-to-noise ratio}

\acrodef{LRT}{likelihood ratio test}

\acrodef{GLRT}{generalized likelihood ratio test}

\acrodef{GML}{generalized maximum likelihood}

\acrodef{LOS}{line-of-sight}

\acrodef{NLOS}{non-line-of-sight}

\acrodef{GDOP}{geometric dilution of precision}

\acrodef{GPS}{Global Positioning System}

\acrodef{FIM}{Fisher information matrix}

\acrodef{PEB}{position error bound}

\acrodef{WSN}{Wireless Sensor Network}

\acrodef{MAC}{medium access control}

\acrodef{RSS}{received signal strength}

\acrodef{RTT}{round-trip time}

\acrodef{MIMO}{multiple-input multiple-output}

\acrodef{MF}{matched filter}

\acrodef{ED}{energy detector}

\acrodef{ML}{maximum likelihood}

\acrodef{NL}{nonlinear}

\acrodef{MSE}{mean square error}

\acrodef{RMSE}{root mean square error}

\acrodef{ppm}{part-per-million}

\acrodef{PRP}{pulse repetition period}

\acrodef{ACK}{acknowledge}

\acrodef{UWB}{ultrawide bandwidth}

\acrodef{TNR}{threshold-to-noise ratio}

\acrodef{NLOS}{non line-of-sight}

\acrodef{LOS}{line-of-sight}

\acrodef{LS}{least squares}

\acrodef{IR-UWB}{impulse radio UWB}

\acrodef{FCC}{Federal Communications Commission}

\acrodef{TH}{time-hopping}

\acrodef{PPM}{pulse position modulation}

\acrodef{PAM}{pulse amplitude modulation}

\acrodef{MUI}{multi-user interference}

\acrodef{PDP}{power delay profile}

\acrodef{PPP}{Poisson point process}

\acrodef{DS}{delay spread}

\acrodef{CED}{channel excess delay}

\acrodef{BPZF}{band-pass zonal filter}

\acrodef{SIR}{signal-to-interference ratio}

\acrodef{RFID}{radio frequency identification}

\acrodef{WPAN}{wireless personal area networks}

\acrodef{WWLB}{Weiss-Weinstein lower bound}

\acrodef{DP}{direct path}

\acrodef{MF}{matched filter}

\acrodef{MMSE}{minimum-mean-square-error}

\acrodef{SBS}{serial backward search}

\acrodef{NBI}{narrowband interference}

\acrodef{WBI}{wideband interference}

\acrodef{INR}{interference-to-noise ratio}

\acrodef{CIR}{channel impulse response}

\acrodef{ISI}{inter-symbol interference}

\acrodef{CPR}{channel pulse response}

\acrodef{LRT}{likelihood ratio test}

\acrodef{RADAR}{RADAR}

\acrodef{MUR}{Multistatic RADAR}

\acrodef{MUI}{multi-user interference}

\acrodef{EM}{electromagnetic}

\acrodef{CW}{continuous wave}

\acrodef{RF}{radiofrequency}

\acrodef{FCC}{Federal Communications Commission}

\acrodef{EIRP}{effective radiated isotropic power}

\acrodef{RCS}{radar cross section}

\acrodef{BAV}{balanced antipodal Vivaldi}

\acrodef{PRake}{partial Rake}

\acrodef{RTLS}{real time locating system}

\acrodef{CRB}{Cram\'{e}r-Rao bound}

\acrodef{ZZB}{Ziv-Zakai bound}

\acrodef{TOA}{time-of-arrival}

\acrodef{TOF}{time-of-flight}

\acrodef{WSN}{wireless sensor network}

\acrodef{MAC}{medium access control}

\acrodef{RSS}{received signal strength}

\acrodef{TDOA}{time difference-of-arrival}

\acrodef{RF}{radiofrequency}

\acrodef{RTT}{round-trip time}

\acrodef{AOA}{angle-of-arrival}

\acrodef{MF}{matched filter}

\acrodef{ED}{energy detector}

\acrodef{ML}{maximum likelihood}

\acrodef{MUR}{Multistatic radar}

\acrodef{HDSA}{high-definition situation-aware}

\acrodef{RRC}{root raised cosine}

\acrodef{OFDM}{orthogonal frequency division multiplexing}

\acrodef{IF}{intermediate frequency}

\acrodef{PHY}{physical layer}

\acrodef{S-V}{Saleh-Valenzuela}

\acrodef{UHF}{ultra-high frequency}

\acrodef{PR}{pseudo-random}

\acrodef{SoC}{System on Chip}

\acrodef{SoP}{System on Package}

\acrodef{SPMF}{Single-Path Matched Filter}

\acrodef{IMF}{Ideal Matched Filter}

\acrodef{SCR}{signal-to-clutter ratio}

\acrodef{BEP}{bit error probability}

\acrodef{BER}{bit error rate}

\acrodef{WSR}{wireless sensor radar}

\acrodef{HPBW}{half power beam width}

\acrodef{LEO}{localization error outage}

\acrodef{WSS}{wide-sense stationary}

\acrodef{TR}{time-reversal}

\acrodef{WSSUS}{WSS with uncorrelated scattering}

\acrodef{GP}{Gaussian process}

\acrodef{IMU}{inertial measurement unit}

\newcommand{\fourier}[1]{\mathcal{F} \left [ #1\right ]}
\newcommand{\invfourier}[1]{\mathcal{F}^{-1} \left [ #1\right ]}

\newcommand{\boldA} {{\bf{A}}}

\newcommand{\bolds} {{\bf{s}}}
\newcommand{\boldf} {{\bf{f}}}
\newcommand{\bolda} {{\bf{a}}}
\newcommand{\boldb} {{\bf{b}}}
\newcommand{\boldp} {{\bf{p}}}
\newcommand{\bolde} {{\bf{e}}}
\newcommand{\boldk} {{\bf{k}}}

\newcommand{\boldX} {{\bf{X}}}

\newcommand{\boldT} {{\bf{T}}}

\newcommand{\boldH} {{\bf{H}}}

\newcommand{\boldG} {{\bf{G}}}

\newcommand{\boldD} {{\bf{D}}}

\newcommand{\boldI} {{\bf{I}}}
\newcommand{\boldr} {{\bf{r}}}

\newcommand{\boldh} {{\bf{h}}}

\newcommand{\rect}[1] {\text{Rect} \left ({#1} \right )}
\newcommand{\sinc}[1] {\text{Sinc} \left ({#1} \right )}

\newcommand{\diag}[1]{{\rm diag} \left ( #1 \right )}

\newcommand{\versorr} {\hat{\bf{r}}}
\newcommand{\versorp} {\hat{\bf{p}}}

\newcommand{\versora} {\hat{\bf{a}}}

\newcommand{\versorx} {{\hat{\bf{x}}}}
\newcommand{\versory} {{\hat{\bf{y}}}}
\newcommand{\versorz} {{\hat{\bf{z}}}}

\newcommand{\Scal} {\mathcal{S}}

\newcommand{\Lx} {L_{\text{x}}}
\newcommand{\Ly} {L_{\text{y}}}

\newcommand{\Nm} {N^{(m)} }
\newcommand{\Ni} {N^{(i)} }

\newcommand{\np} {n_{\text{p}}}
\newcommand{\ns} {n_{\text{s}}}
\newcommand{\up} {u_{\text{p}}}

\newcommand{\wavefreq} {{\boldsymbol{\kappa}}}
\newcommand{\wavefreqx} {\kappa_{x}}
\newcommand{\wavefreqy} {\kappa_{y}}
\newcommand{\wavefreqz} {\kappa_{z}}
\newcommand{\wavefreqp} {\overline{\wavefreq}}
\newcommand{\wavefreqxp} {\overline{\kappa}_x}
\newcommand{\wavefreqyp} {\overline{\kappa}_y}
\newcommand{\wavefreqzp} {\overline{\kappa}_z}

\newcommand{\wavefreqbis} {\boldk}
\newcommand{\wavefreqbisx} {k_x}
\newcommand{\wavefreqbisy} {k_y}

\newcommand{\kappazero} {k_0}

\newcommand{\EMsymb} {\mathsf}

\newcommand{\Green} {{ G_0} }
\newcommand{\Greent} {{\EMsymb {{\underline{G}}}}  }
\newcommand{\Greenej} {{\EMsymb {{\underline{G}}}_{\text{EJ}}}  }
\newcommand{\Greenem} {{\EMsymb {{\underline{G}}}_{\text{EM}}}  }
\newcommand{\Greenhj} {{\EMsymb {{\underline{G}}}_{\text{HJ}}}  }
\newcommand{\Greenhm} {{\EMsymb {{\underline{G}}}_{\text{HM}}}  }

\newcommand{\TildeGreen} {{ \widetilde{ G}_0} }
\newcommand{\TildeGreenTot} {\underline{ \widetilde{\cal{H}} }}
\newcommand{\TildeGreenTotej} {\underline{ \widetilde{\cal H}}_{\text{EJ}} }

\newcommand{\TildeGreenTotejxx} {{ \widetilde{\cal H}}_{\text{EJ}}^{(xx)} }

\newcommand{\Dm} {{\EMsymb {{\underline{D}}}}  }

\newcommand{\Dmm} {{\EMsymb {{\underline{D}}}^{\, (m)}}  }
\newcommand{\Dmxx} {{\EMsymb {{{D}}}}^{(xx)}  }
\newcommand{\Dmxy} {{\EMsymb {{{D}}}}^{(xy)}  }
\newcommand{\Dmxz} {{\EMsymb {{{D}}}}^{(xz)}  }
\newcommand{\Dmyx} {{\EMsymb {{{D}}}}^{(yx)}  }
\newcommand{\Dmyy} {{\EMsymb {{{D}}}}^{(yy)}  }
\newcommand{\Dmyz} {{\EMsymb {{{D}}}}^{(yz)}  }
\newcommand{\Dmzx} {{\EMsymb {{{D}}}}^{(zx)}  }
\newcommand{\Dmzy} {{\EMsymb {{{D}}}}^{(zy)}  }
\newcommand{\Dmzz} {{\EMsymb {{{D}}}}^{(zz)}  }

\newcommand{\Dmje} {{\EMsymb {{\underline{D}}}_{\text{JE}}^{\, (m)}}  }
\newcommand{\Dmjh} {{\EMsymb {{\underline{D}}}_{\text{JH}}^{\, (m)}} }
\newcommand{\Dmme} {{\EMsymb {{\underline{D}}}_{\text{ME}}^{\, (m)}}  }
\newcommand{\Dmmh} {{\EMsymb {{\underline{D}}}_{\text{MH}}^{\, (m)}}  }

\newcommand{\Dw} {{\EMsymb {{\underline{D}}}_{\text{w}}}  }
\newcommand{\DR} {{\EMsymb {{\underline{R}}} }  }

\newcommand{\Em} {{\EMsymb E}}
\newcommand{\Hm} {{\EMsymb H}}

\newcommand{\Emmt} {\EMsymb E^{(m)}_{\text{t}}}
\newcommand{\Hmmt} {\EMsymb H^{(m)}_{\text{t}}}
\newcommand{\Emt} {\EMsymb E_{\text{t}}}
\newcommand{\Hmt} {\EMsymb H_{\text{t}}}

\newcommand{\Am} {\EMsymb A}
\newcommand{\Bm} {\EMsymb B}
\newcommand{\Zm} {\EMsymb Z}

\newcommand{\TildeEm} { \widetilde{\EMsymb E}}
\newcommand{\TildeHm} {\widetilde{\EMsymb H}}
\newcommand{\TildeEmt} { \widetilde{\EMsymb E}_{\text{t}}}
\newcommand{\TildeHmt} {\widetilde{\EMsymb H}_{\text{t}}}
\newcommand{\TildeAm} { \widetilde{\EMsymb A}}
\newcommand{\TildeDm} {\underline{ \widetilde{\EMsymb D}}}

\newcommand{\Jm} {{\EMsymb J}}
\newcommand{\Mm} {{\EMsymb M}}
\newcommand{\Jmi} {\EMsymb J_{\text{imp}}}
\newcommand{\Jms} {\EMsymb J_{\text{s}}}
\newcommand{\Mmi} {\EMsymb M_{\text{imp}}}
\newcommand{\Mms} {\EMsymb M_{\text{s}}}

\newcommand{\TildeJm} {{ \widetilde{\EMsymb J}}}
\newcommand{\TildeJmi} {{ \widetilde{\EMsymb J}}_{\text{imp}}}
\newcommand{\TildeJms} {{ \widetilde{\EMsymb J}}_{\text{s}}}
\newcommand{\TildeMm} {{ \widetilde{\EMsymb M}}}

\newcommand{\TildeMms} {{ \widetilde{\EMsymb M}}_{\text{s}}}

\newcommand{\Htf} {\underline{{ \widetilde{\mathcal H}}}}
\newcommand{\Htfxx} {{{ \widetilde{\mathcal H}}}^{(xx)}}

\newcommand{\Ym} {{\EMsymb Y}}
\newcommand{\Yje} {{\EMsymb Y}_{\text{JE}}}
\newcommand{\Yjh} {{\EMsymb Y}_{\text{JH}}}
\newcommand{\Yme} {{\EMsymb Y}_{\text{ME}}}
\newcommand{\Ymh} {{\EMsymb Y}_{\text{MH}}}

\newcommand{\Phim} {{\mathbf{\Phi}}}
\newcommand{\TildePhim} {\widetilde{\mathbf{\Phi}}}
\newcommand{\OPhim} {\breve{\mathbf{\Phi}}}
\newcommand{\Ophim} {\tilde{\phi}} 

\newcommand{\boldzero} {\boldsymbol{0}}

\newcommand{\curl} {\nabla  {\scriptstyle \mathsf{x}}\, }
\newcommand{\scalprod} {\boldsymbol{\cdot}}

\newcommand{\innerprod}[2] {\left < {#1}\, , {#2} \right >}
\newcommand{\crossprod}[2] {{#1}\, { \scriptstyle \mathsf{x}}\, {#2} }

\newcommand{\inttwo} {\int_{{\cal{R}}^2}}
\newcommand{\intthree} {\int_{{\cal{R}}^3}}

\newcommand{\ajmn} {a_{J_n}^{(m)}}
\newcommand{\bjmn} {b_{J_n}^{(m)}}
\newcommand{\ammn} {a_{M_n}^{(m)}}
\newcommand{\bmmn} {b_{M_n}^{(m)}}
\newcommand{\ajiu} {a_{J_u}^{(i)}}
\newcommand{\bjiu} {b_{J_u}^{(i)}}
\newcommand{\amiu} {a_{M_u}^{(i)}}
\newcommand{\bmiu} {b_{M_u}^{(i)}}
\newcommand{\ajm} {\bolda_J^{(m)}}
\newcommand{\bjm} {\boldb_J^{(m)}}
\newcommand{\amm} {\bolda_M^{(m)}}
\newcommand{\bmm} {\boldb_M^{(m)}}
\newcommand{\amv} {\bolda^{(m)}}
\newcommand{\bmv} {\boldb^{(m)}}
\newcommand{\fmv} {\boldf^{(m)}}
\newcommand{\fMv} {\boldf^{(M)}}
\newcommand{\aiv} {\bolda^{(i)}}
\newcommand{\aonev} {\bolda^{(1)}}
\newcommand{\biv} {\boldb^{(i)}}
\newcommand{\fiv} {\boldf^{(i)}}

\newcommand{\emv} {\bolde^{(m)}}
\newcommand{\hmv} {\boldh^{(m)}}
\newcommand{\eMv} {\bolde^{(M)}}

\newcommand{\CalP} {{\cal{P}}}

\begin{document}
\title{Reconfigurable Electromagnetic Environments: A General Framework}

\author{
\IEEEauthorblockN{Davide~Dardari,~\IEEEmembership{Senior~Member,~IEEE}}
\IEEEcompsocitemizethanks{\IEEEcompsocthanksitem 
 D.~Dardari is with the 
   Dipartimento di Ingegneria dell'Energia Elettrica e dell'Informazione ``Guglielmo Marconi"  (DEI), WiLAB-CNIT, 
   University of Bologna, Cesena Campus, 
   Cesena (FC), Italy, (e-mail: davide.dardari@unibo.it). 
    }
}

\maketitle

\begin{abstract}
The recent introduction of the \acp{SRE} paradigm, enabled by \acp{RIS}, has put in evidence the need for physically-consistent models and design tools for communication systems integrating \ac{EM} and signal processing theories. In this perspective, starting from rigorous \ac{EM} arguments, in this paper we propose a general framework for the characterization and design of programmable \ac{EM} environments. We first show that any linear \ac{EM} environment in the presence of boundary conditions can be interpreted as a  space-variant linear feedback filter. Then we provide a methodology to characterize programmable \ac{EM} systems as a linear graph described by matrix operators thus leading to the determination of the transfer function of the \ac{EM} system. 
Finally, some examples are given related to the characterization and design of \acp{RIS},  also showing that some previous results in the literature are just particular cases of our general framework.
\end{abstract}



\section{Introduction}

Recently, the \acf{SRE} concept has been introduced as one of the new design paradigms of next-generation networks \cite{DiRZapDebAloYueRosTre:20,BarHamLonMonRamVelAleBil:22}. 
While in current communication systems the propagation environment is considered as given and the communicating devices are optimized to adapt to it, thanks to the deployment of programmable \ac{EM} devices such as \acfp{RIS}, in \acp{SRE} the environment enters into the design and optimization loop. This is expected to pave the way to more flexible wireless networks offering improved performance in terms of achievable rate, interference shaping, coverage extension, energy, and complexity reduction. 

An extensive research activity has been devoted to the study of \ac{RIS}-aided communication and localization systems, for example, the papers  \cite{DarDec:J21,BjoWymMatPopSanCar:22,AbrDarDiR:J21}. 
 At the same time, the introduction of  extremely electrically large antennas made of metasurfaces working at high frequency opened the door to the exploitation of the radiating near-field characteristics of the radio channel even at practical distances \cite{JenWal:08,Miller:19,DarDec:J21,ZhaShlGuiDarImaEld:J22,DecDar:J21}.
Despite the wide literature available today on the subject, the main shortcomings of previous studies are summarized in the following. 
On the one hand,  oversimplified but tractable models, often relying on physically-inconsistent assumptions, have been mainly considered in system-level design and optimization that do not capture important peculiarities of the \acp{EMO} composing the system. An example of a typically neglected \ac{EM} phenomenon exhibited in smart surfaces is given by the  Floquet modes, i.e., spurious reflections of surfaces characterized by a periodic impedance that might generate interference at non-desiderate angles and hence compromising the system performance  \cite{DiaTre:21}.  Moreover, the search for the ultimate theoretical limits cannot  rescind from an accurate description of the underlying \ac{EM} phenomena \cite{OzdBjoLar:20,Dar:J20}.
 On the other hand, in the \ac{EM} community, more emphasis has been put mainly on the characterization of single devices, often resorting to extensive \ac{EM}-level simulations that can be barely included in a system-level analysis or real-time system optimization cycle 
 \cite{Tre:15, AchSalCal:15,DegVitDiRTre:22, AsaAlbTcvRubRadTre:16,DiaTre:21,MarMac:22}.
 %
 %
 This dichotomy has made sense until now in a context where the \ac{EMO} (for instance,  the antenna) was seen as a  ``sensor" or ``actuator/transducer" by the communication theorists community. 
In a near future, characterized by the presence of reconfigurable \ac{EM} environments, the main objective will be the optimization of the system response, even in real-time, through the configuration of \ac{EMO}'s parameters  (e.g., the reflection properties of a metasurface-based \ac{RIS}). This requires a holistic system view that calls for physically-consistent models and design tools integrating  signal processing and \ac{EM} theory  \cite{ZhuWanDaiDebPoo:22}.
 Some recent works have undertaken this path. The investigation of the degrees of freedom of the wireless channel when using \acp{LIS}, also dubbed as XL-MIMO or holographic MIMO, by modeling the surface as a continuous of infinitesimal antenna elements, has been conducted in \cite{JenWal:08,Miller:19,Dar:J20,DarDec:J21}.
 Regarding \acp{RIS}, the work  \cite{GraDiR:21} proposes the modeling of a \ac{RIS} as a set of elementary coupled dipoles and then characterizes its response in terms of impedance matrices.  
 In \cite{NajJamSchPoo:21}, continuous and discrete models for  the response   of a perfect reflecting tile (i.e., a subset of the \ac{RIS}) are derived  starting from \ac{EM} arguments and subsequently used in a two-step optimization approach targeted to achieving the desired communication quality by considering an ensemble of reconfigurable tiles. 
 A recent survey that includes a critical discussion on \ac{RIS} modeling can be found in \cite{DiRDanTre:22}. 
 
Moving more at a system level, the authors in \cite{PooBroTse:05} define the concept of  system \ac{EM} transfer function in the wavenumber domain similarly to what is done in the signal processing community with the classical transfer function of a linear system  in the frequency domain. This concept has been     further elaborated in recent papers \cite{PizSanMar:22,PizSanMar:22a,BjoWymMatPopSanCar:22,PizLozRanMar:23}. In particular, \cite{PizSanMar:22} and \cite{PizSanMar:22a} generalize the concept of \ac{EM} transfer function for a  stochastic Gaussian propagation environment characterized by rich scattering.
Paper \cite{PizLozRanMar:23} derives the \ac{EM} transfer function in the particular case of an infinite size surface with constant reflection coefficient by establishing a connection between the well-known image theorem in \ac{EM} theory and the \ac{EM} transfer function. 
Finally,  a generalized expression of the \ac{EM} field reflected by
an \ac{EM} skin in far-field and radiative near-field regimes is given in \cite{OliSalMas:22} and subsequently used to derive a unified method for the design of anomalous-reflecting and focusing \ac{EM} skins. In \cite{MasBenDaRGouBaoOliPolRocSal:21}, the same authors propose a numerical optimization problem for the design of \acp{SRE} aiming at synthesizing a desired \ac{EM} field distribution over a target region. 
 
Previously mentioned studies are limited to the definition of the \ac{EM} transfer function as a ``black box" without providing any general methodology to compute it, apart from a few oversimplified scenarios or under oversimplified assumptions (e.g., no polarization, infinite uniform surfaces, point-wise scatterers, far-field regime). Moreover, none of the papers address the characterization of the \ac{EM} transfer function for reconfigurable \acp{EMO} such as \acp{RIS}.
 
To fill the gap, in this paper we propose a general framework, starting from rigorous \ac{EM} arguments valid both in far-field and (radiative and reactive) near-field regimes, aimed at giving  a system-theoretic and physically-consistent interpretation of reconfigurable \ac{EM} environments. It is shown that any system involving linear \acp{EMO} can be described as a space-variant feedback system. Specifically, we show that boundary conditions applied to a generic \ac{EMO}  translate into a feedback system representation. 
Since the analysis and design of  space-variant feedback systems are generally complex, inspired by mode-matching techniques, we propose an approach involving linear algebra operations over a graph that is useful to  design and characterize \ac{EM} systems in the presence of reconfigurable \acp{EMO}. Subsequently, we derive the relationship between the \ac{EM}  transfer function and the linear algebra representation of the system, with particular emphasis on the characterization   of smart surfaces.  
To illustrate the use of the proposed framework, some examples of  the derivation of the \ac{EM}  transfer function  are given. Some of them  show  that  previous results in the literature can be seen as  particular cases of our general framework. A final example is given targeted to the design of a \ac{RIS} minimizing spurious interference caused by the presence of Floquet modes and/or other \ac{EM} sources.

\subsection{Paper Organization}

The rest of the paper is organized as follows: 
Sec.~\ref{Sec:Problem} formulates the problem of the description of a reconfigurable \ac{EM} system composed of a certain number of \acp{EMO}.
In Secs.~\ref{Sec:Linear}, \ref{Sec:Basis}, and \ref{Sec:Coupling}, a linear algebra description of the \ac{EM} system is developed  through  the introduction of harmonic basis functions (\emph{modes}) (Sec.~\ref{Sec:Basis}) and, subsequently,    the derivation of the expressions for modes coupling (Sec.~\ref{Sec:Coupling}), thus providing a graph-based interpretation of the  system. The modeling and characterization of reconfigurable surfaces in terms of boundary conditions are addressed in Sec.~\ref{Sec:D}.   
The relationship between the linear algebra description and the system \ac{EM}  transfer function is explained Sec.~\ref{Sec:Examples},  where some examples are given for some particular cases of interest. Finally,  in Sec.~\ref{Sec:Conclusion} the conclusions are drawn.

\subsection{Notation and Definitions}
Lowercase bold variables denote vectors in the 3D space, i.e.,  $\boldr=\versorx \cdot r_x + \versory \cdot r_y + \versorz \cdot r_z$
 is a vector with cartesian coordinates $(r_x, r_y, r_z)$,
 $\versorr$ is a unit vector denoting its direction, and  $r=|\boldr|$ denotes its magnitude, where $\versorx$, $\versory$, and $\versorz$ represent the unit vectors in the $x$, $y$ and $z$ directions, respectively.  The cross product between vectors $\boldp$ and 
 $\boldr$ is indicated with $\crossprod{\boldp}{\boldr}$, whereas the scalar product with ${\boldp} \scalprod {\boldr}$. $\delta(x)$ and $\delta_n$ represent, respectively, the Dirac delta pseudo-function and its discrete counterpart (Kronecker delta). Multi-variable versions can be defined as well, i.e., $\delta(\boldr)=\delta(x)\, \delta(y)\, \delta(z)$ and $\delta_{n,m}=\delta_{n-m}$. 
Sans sherif capital letters (e.g., $\Em(\boldr)$, $\Jm(\boldr)$) represent \ac{EM} vector functions (in the following named \emph{fields}), whereas scalar functions are denoted with roman lowercase letters, i.e., $\phi(\boldr)$. Boldface capital letters are matrices (e.g., $\boldA$), where $\boldI_N$ is the identity matrix of size $N$, $\boldzero_N$ the zero matrix of size $N$, $a_{n,m}=[\boldA]_{n,m}$ represents the $(n,m)$th element of matrix $\boldA$,  and $^*$ indicates the complex conjugate operator.
$\nabla^2  \, \Em (\boldr)$ and $\curl \Em(\boldr)$ are, respectively, the Laplacian and the curl of the vector function $\Em(\boldr)$.
Surfaces, contours, and volumes  are indicated with calligraphic letters $\Scal$. 
Any linear transformation of a field $\Am(\boldr)$ into a field $\Bm(\boldr)$ can be represented as $\Bm(\boldr)=\Dm \cdot \Am(\boldr)$, where  $\Dm$ is a dyadic that can be expressed as 
\begin{equation} \nonumber
\Dm=\left (
\begin{array}{ccc}
   \Dmxx &  \Dmxy & \Dmxz \\
   \Dmyx &  \Dmyy & \Dmyz \\
    \Dmzx &  \Dmzy & \Dmzz \\ 
     \end{array} 
\right ) \, .
\end{equation}
 Typically, $\Dm$ depends on the position $\boldr$ even though this is not shown explicitly to lighten the notation. Often an operation involving a dyadic can be expressed in matrix form.
 Define $\sinc{x}=\sin(\pi\, x)/(\pi \, x)$ for $x \neq 0$, 1 for $x=0$, and $\rect{x}=1$ for $|x|<1/2$, zero otherwise.  Furthermore, denote by $\mu$, $\epsilon$, and $\eta=\sqrt{\mu/\epsilon}$ the  free-space permittivity, permeability and impedance, respectively, and $c$  the speed of light.   


\begin{figure}[t]
\centerline{\includegraphics[width=0.9\columnwidth]{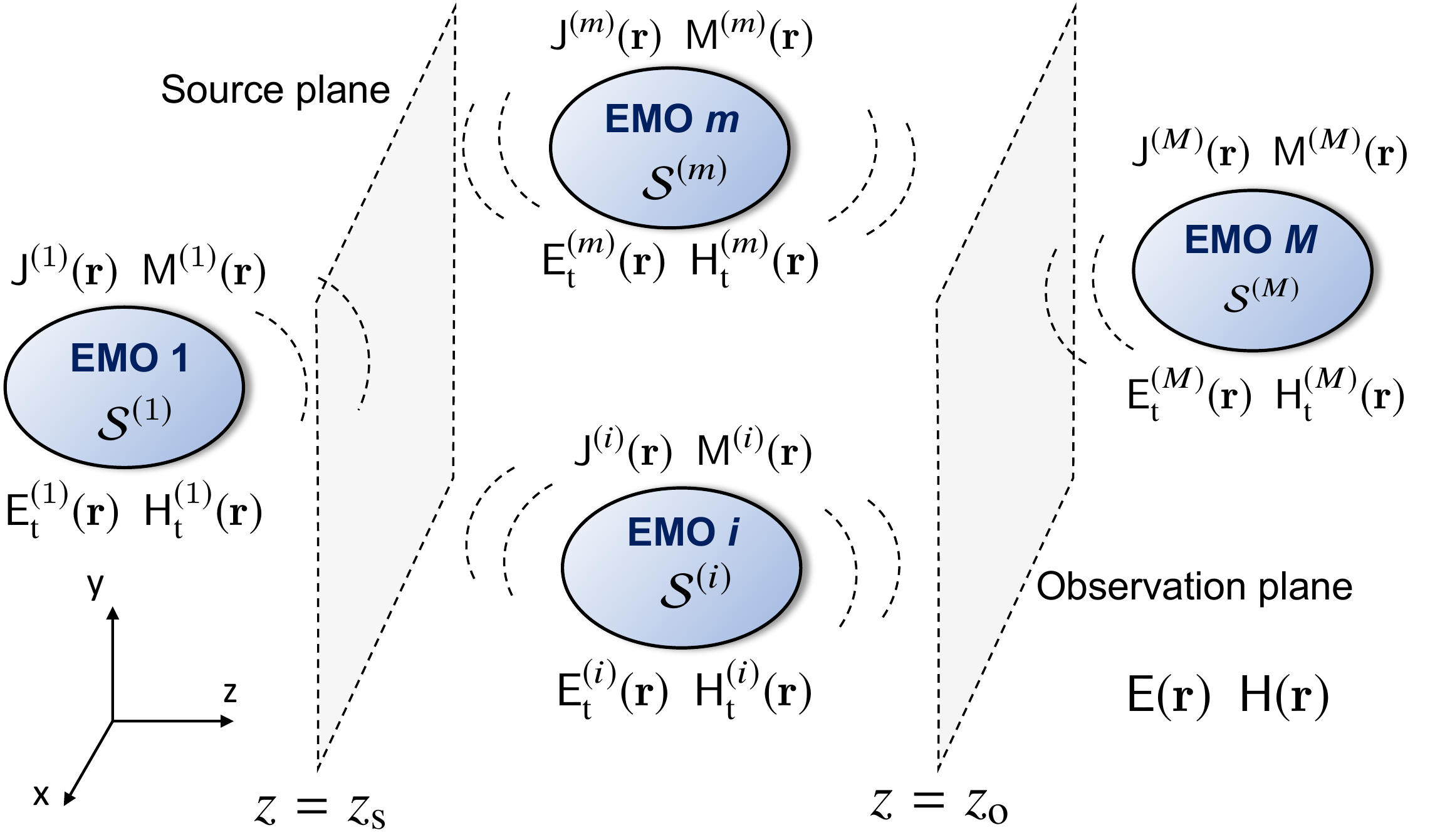}}
\caption{General \ac{EM} scenario with interacting \acp{EMO}. }
\label{Fig:EMScenario}
\end{figure}

\section{Reconfigurable EM System}
\label{Sec:Problem}

\subsection{Problem formulation}

We consider $M$ linear time-invariant \acp{EMO}  are present in the system sketched in Fig. \ref{Fig:EMScenario}, and we indicate with $\Scal^{(m)}$, $m=1, 2, \ldots , M$, the surface that encloses the $m$th \ac{EMO}. Each surface can represent the physical surface of the \ac{EMO}  or any arbitrary surface which encloses it. 
We work in the frequency domain, where time-harmonic excitations $\exp(\jmath \omega t)$ are assumed, being $\omega$ the angular frequency, and all fields and sources are phasors.
Denote with  $\Jm^{(m)} (\boldr)$ and $\Mm^{(m)} (\boldr)$ the electric and magnetic currents densities, respectively, on surface $\Scal^{(m)}$.\footnote{In general, $\Scal^{(m)}$ may be a surface or a contour (e.g., a wire antenna). With a little abuse of nomenclature, for further convenience, we will still denote it as \emph{surface}. 
} 
 Notice that $\Jm^{(m)}(\boldr)$ and $\Mm^{(m)}(\boldr)$ are zero outside $\Scal^{(m)}$.
 The total electric and magnetic currents present in the system are given by $\Jm(\boldr)=\sum_{m=1}^M \Jm^{(m)}(\boldr)$ and $\Mm(\boldr)=\sum_{m=1}^M \Mm^{(m)}(\boldr)$, respectively.
In general,  $\Jm^{(m)} (\boldr)$ and $\Mm^{(m)} (\boldr)$ can be decomposed into the sum of the impressed $\left ( \Jmi^{(m)}(\boldr)\, , \Mmi^{(m)}(\boldr) \right )$   (if any) and induced $\left( \Jms^{(m)}(\boldr)\, , \Mms^{(m)}(\boldr)\right ) $ currents
\begin{align} \label{eq:JmMn}
\Jm^{(m)}(\boldr) & = \Jms^{(m)}(\boldr)+ \Jmi^{(m)}(\boldr) 
&  \Mm^{(m)}(\boldr) & = \Mms^{(m)}(\boldr)+ \Mmi^{(m)}(\boldr) \, .
\end{align}

According to the equivalent principle \cite{BalB:16}, the introduction of (fictitious)  induced currents, satisfying the boundary conditions at the corresponding surface, permits to consider the induced current sources to radiate into an unbounded space. 
Therefore,  the \ac{EM} field, i.e., the electric and magnetic fields,  generated by all the currents present in the system at the generic location $\boldr$ can be computed under the free-space condition, that is, 
\begin{align}  \label{eq:EmHm}
\left (
\begin{array}{c}
   \Em(\boldr)   \\
    \Hm(\boldr)   \\  
     \end{array} 
\right )
= 
\left (
\begin{array}{cc}
   \Greenej   &  \Greenem  \\
   \Greenhm & \Greenhj  \\  
     \end{array} 
\right )
\cdot 
\left (
\begin{array}{c}
   \Jm(\boldr) \\
    \Mm(\boldr)  \\  
     \end{array} 
\right )
= 
\Greent \cdot
\left (
\begin{array}{c}
   \Jm(\boldr) \\
    \Mm(\boldr)  \\  
     \end{array} 
\right ) 
\end{align}
where the above dyadics are given by \cite{HarringtonBook:2001}
\begin{align} \label{eq:gej}
\Greenej  \cdot \Jm(\boldr)   &= \frac{1}{\jmath \omega \epsilon}  \curl \curl \int_{\Scal} \Green (\boldr - \bolds) \, \Jm(\bolds) \, d\bolds   \\
\label{eq:gem}
\Greenem \cdot \Mm(\boldr)  &=- \curl  \int_{\Scal} \Green (\boldr - \bolds) \, \Mm(\bolds) \, d\bolds \\
\label{eq:ghj}
\Greenhj \cdot \Jm(\boldr)   &= \curl  \int_{\Scal} \Green (\boldr - \bolds) \, \Jm(\bolds) \, d\bolds  \\
\label{eq:ghm}
\Greenhm \cdot \Mm(\boldr)  &= \frac{1}{\jmath \omega \mu}  \curl \curl \int_{\Scal} \Green (\boldr - \bolds) \, \Mm(\bolds) \, d\bolds  
\end{align}
 being $\Scal = \bigcup_{m=1}^M \Scal^{(m)}$. The function 
\begin{equation}
	\Green(\boldr)=\frac{\exp (-\jmath \kappazero |\boldr| ) }{4 \pi |\boldr|} 
\end{equation}
is the free-space scalar Green function, where  $\kappazero=2\pi / \lambda$ is the wavenumber and $\lambda=2\pi c/\omega$ is the wavelength. It can be easily noticed from the previous equations that the propagation phenomenon operates as a space-invariant linear filter because $\Green(\boldr)$ appears in \eqref{eq:gej}-\eqref{eq:ghm} as a  function of only the difference $\boldr - \bolds$.   

\begin{figure}[t]
\centerline{\includegraphics[width=0.6\columnwidth]{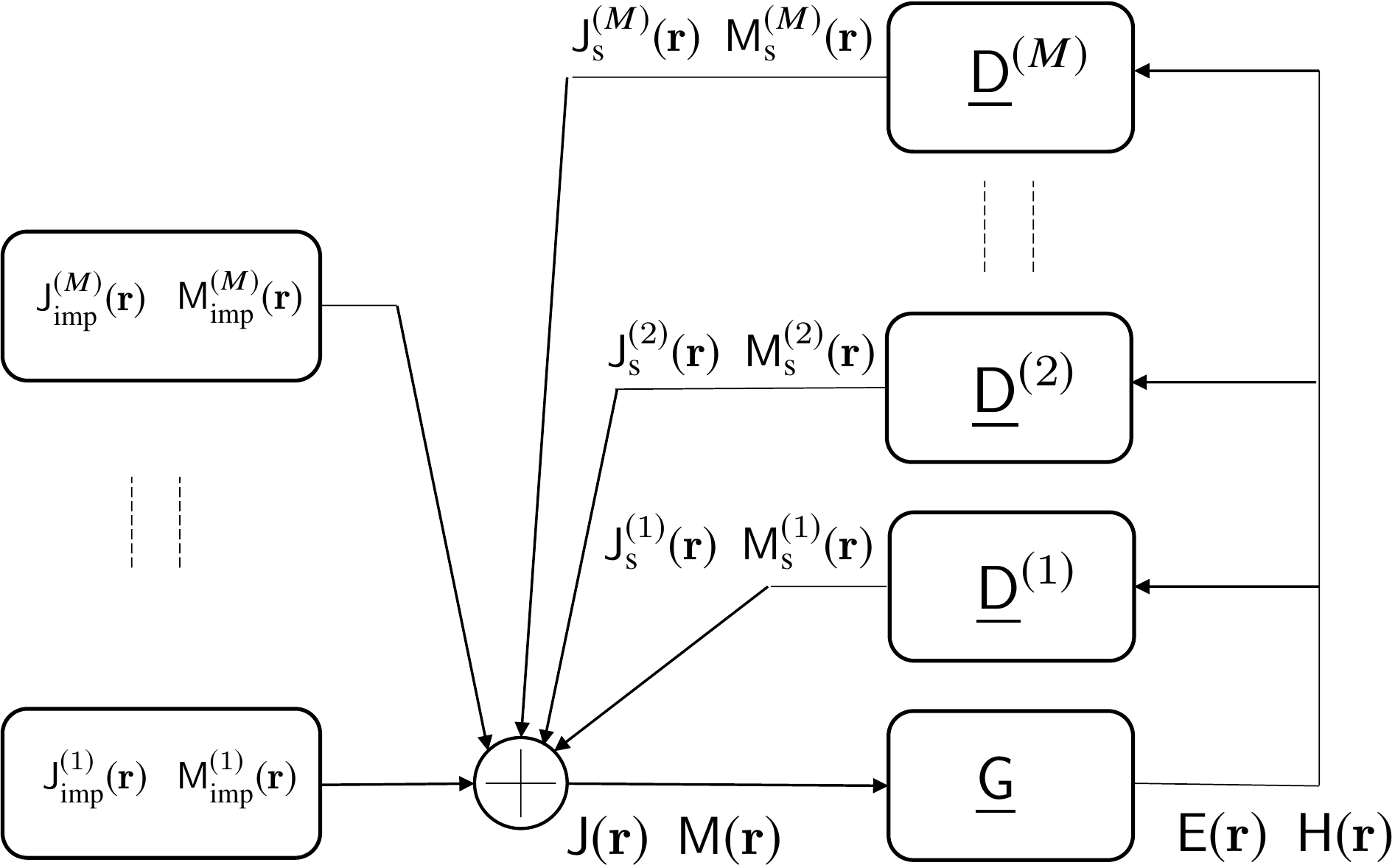}}
\caption{\ac{EM} scenario as a  space-variant feedback system. }
\label{Fig:EMSystem}
\end{figure}

Define the \ac{EM} field $\left ( \Emt^{(m)}(\boldr) \, , \Hmt^{(m)}(\boldr) \right )$ tangent to surface $\Scal^{(m)}$ of the $m$th \ac{EMO}.  
For any linear time-invariant \ac{EMO}, the induced  currents are linear functionals of the \ac{EM} field tangent to the surface so that they can be described  as follows (\emph{constitutive equation}) \cite{Miller:19} 
\begin{align} \label{eq:JMs}
\left (
\begin{array}{c}
   \Jms^{(m)}(\boldr)   \\
    \Mms^{(m)}(\boldr)   \\  
     \end{array} 
\right )
= 
\left (
\begin{array}{cc}
   \Dmje &  \Dmjh  \\
   \Dmme & \Dmmh  \\  
     \end{array} 
\right )
\cdot 
\left (
\begin{array}{c}
   \Emt^{(m)}(\boldr) \\
    \Hmt^{(m)}(\boldr) \\  
     \end{array} 
\right )
= 
\Dmm  \cdot
\left (
\begin{array}{c}
   \Emt^{(m)}(\boldr) \\
    \Hmt^{(m)}(\boldr) \\  
     \end{array} 
\right )
\end{align}
where the dyadic $\Dmje$ can be expressed in general as
\begin{align} \label{eq:Dmje1}
\Dmje \cdot  \Emt^{(m)}(\boldr)  &=  \int_{\Scal^{(m)}} \Dm_{\text{JE}}^{\, (m)}(\boldr, \bolds) \, \Emt^{(m)}(\bolds) \, d\bolds   
\end{align}
being $\Dm_{\text{JE}}^{\, (m)}(\boldr, \bolds)$ the impulse response dyadic describing completely the relationship between the electric field and the induced current at the $m$th  \ac{EMO}.  Similar expressions can be written for the dyadics $\Dmjh$,  $\Dmme$, and $\Dmmh$. For instance, in non-magnetic scatterers, the induced magnetic current is zero, i.e.,  $\Mms^{(m)}(\boldr)=0$, then $ \Dmme = \Dmmh  =0$ \cite{ChenB:18}. The  particular form of $\Dmm$ depends on the model adopted for the \ac{EMO} and the target level of accuracy. Some examples are provided in Sec. \ref{Sec:Examples}.
The previous relationships can be represented in the scheme of Fig. \ref{Fig:EMSystem} where it is evident that  any \ac{EM} scenario can be viewed as a feedback system in which the impressed currents represent the inputs, the propagation phenomenon $\Greent$ is a space-invariant filter and each \ac{EMO} can be seen as a space-variant filter (in analogy with time-variant filters) whose output consists of the induced currents. 
The above feedback system involves integral equations whose solution is, in general, a complex problem typically addressed numerically. In this paper, we introduce a methodology to bypass this problem.

\subsection{Wavenumber Domain Representation}

For what follows, it is convenient to introduce the representation of the fields in the wavenumber domain $\wavefreq = \wavefreqx \versorx + \wavefreqy \versory + \wavefreqz \versorz$ through the 3D Fourier transform. Specifically, given a generic field $\Am(\boldr)$, we can write 
\begin{align}
  \TildeAm (\wavefreq) &=\fourier{\Am(\boldr)}= \intthree \Am(\boldr) \, e^{-\jmath \, \wavefreq \scalprod \boldr} \, d \boldr
  \end{align}
\begin{align} 
 \Am(\boldr) &= \invfourier{\TildeAm (\wavefreq)}= \frac{1}{(2 \pi)^3}  \intthree  \TildeAm (\wavefreq)  \, e^{\jmath \, \wavefreq \scalprod \boldr} \, d \wavefreq \, .
 \label{eq:invfourier}
\end{align}

When applied to the \ac{EM} field,  the inverse Fourier representation in  \eqref{eq:invfourier} expresses the \ac{EM} field in terms of mathematical plane waves \cite{CleB:96}.  By inspection of \eqref{eq:invfourier}, the plane wave with wavenumber $\wavefreq$,  travels in the positive $w$-direction  (forward wave), with $w \in \{ x,y,z  \}$, when its component $\wavefreq_{w}<0$,  whereas when $\wavefreq_{w}>0$ the wave travels in the negative $w$-direction (backward wave).   It follows that 
\begin{equation} \label{eq:Greenk}
  \TildeGreen(\wavefreq)=\fourier{\Green(\boldr)}= \frac{1}{|\wavefreq|^2- \kappazero^2} \
\end{equation}
and from \eqref{eq:gej}-\eqref{eq:ghm}, that
%
\begin{align} \label{eq:Emk}
\TildeEm(\wavefreq) &= \fourier{\Em(\boldr)} =\frac{ \jmath \, \TildeGreen(\wavefreq)  \, }{ \omega \epsilon} \crossprod{\wavefreq}{\crossprod{\wavefreq}{ \TildeJm(\wavefreq) }}-\jmath \,  \TildeGreen(\wavefreq) \, \crossprod{\wavefreq}{\TildeMm(\wavefreq)}=\jmath \, \TildeGreen(\wavefreq) \crossprod{\wavefreq}{\left [ \frac{\eta}{\kappazero} {\crossprod{\wavefreq}{  \TildeJm(\wavefreq) }} -  \TildeMm(\wavefreq)\right ]}\\
\TildeHm(\wavefreq) &=  \fourier{\Hm(\boldr)}= \frac{\jmath \, \TildeGreen(\wavefreq)  \, }{ \omega \mu} \crossprod{\wavefreq}{\crossprod{\wavefreq}{  \TildeMm(\wavefreq) }}
+ \jmath \, \TildeGreen(\wavefreq)  \, \crossprod{\wavefreq}{ \TildeJm(\wavefreq)} =\jmath\, \TildeGreen(\wavefreq)  \, 
\crossprod{\wavefreq}{\left [\frac{1}{\kappazero\, \eta} \crossprod{\wavefreq}{\TildeMm(\wavefreq)}   + \TildeJm(\wavefreq) \right ]}
\end{align}
where $\TildeJm(\wavefreq)=\fourier{\Jm(\boldr)}$, $\TildeMm(\wavefreq)=\fourier{\Mm(\boldr)}$, and we have considered that $\fourier{\curl \Am}=\jmath \, \crossprod{\wavefreq}{\TildeAm(\wavefreq)}$. 
 
The spatial filtering operated by the Green operator is evident in \eqref{eq:Greenk} which corresponds to a low-pass filter with a cut-off frequency equal to $\kappazero$. This means that the \ac{EM} field has a spatial low-pass characteristic.  
Incidentally, by defining $\wavefreq_{\boldr}=\kappazero \, \versorr$, the \ac{EM} field at location $\boldr$ in far-field conditions is proportional to the Fourier transform of the sources \cite{BalB:16}, that is,    
\begin{align}
\Em(\boldr) &\simeq \jmath \kappazero   \frac{e^{-\jmath \, \kappazero \, |\boldr|}}{4 \pi |\boldr|}   \,  \crossprod{\versorr}{\left [ \eta \, {\crossprod{\versorr}{  \TildeJm(\wavefreq_{\boldr}) }} +  \TildeMm(\wavefreq_{\boldr})\right ]}  \nonumber \\
\Hm(\boldr) &\simeq \jmath \kappazero   \frac{e^{-\jmath \, \kappazero \, |\boldr|}}{4 \pi |\boldr|}   \,  \crossprod{\versorr}{\left [ \frac{1}{\eta} \, {\crossprod{\versorr}{  \TildeMm(\wavefreq_{\boldr}) }} -  \TildeJm(\wavefreq_{\boldr})\right ]} \, .
\end{align}

By applying the Fourier transform to \eqref{eq:JMs}  we obtain
\begin{align}
\TildeJms^{(m)}(\wavefreq)= \frac{1}{(2\pi)^3} \intthree \TildeDm_{\text{JE}}^{(m)}(\wavefreq,\wavefreqp)  \, \TildeEmt(\wavefreqp)  \, d \wavefreqp + \frac{1}{(2\pi)^3} \intthree \TildeDm_{\text{JH}}^{(m)}(\wavefreq,\wavefreqp)  \, \TildeHmt(\wavefreqp)  \, d \wavefreqp 
\end{align}
where $\TildeDm_{\text{JE}}^{(m)}(\wavefreq,\wavefreqp)$ and $\TildeDm_{\text{JH}}^{(m)}(\wavefreq,\wavefreqp)$ take the role of the \emph{bi-frequency system functions}, or \emph{mapping functions}, in analogy with the bi-frequency system function of time-variant systems \cite{Zad:50}.
They give the induced current response of the \ac{EMO}  at wavenumber $\wavefreq$ when a plane wave with wavenumber $\wavefreqp$ is applied at the input.  Similar expressions hold for $\TildeMms^{(m)}(\wavefreq)$. 
 $\TildeDm_{\text{JE}}^{(m)}(\wavefreq,\wavefreqp)$, $\TildeDm_{\text{JH}}^{(m)}(\wavefreq,\wavefreqp)$, $\TildeDm_{\text{ME}}^{(m)}(\wavefreq,\wavefreqp)$, and $\TildeDm_{\text{MH}}^{(m)}(\wavefreq,\wavefreqp)$,   depend on \ac{EMO}'s characteristics and configuration and might represent the optimization objective in a reconfigurable \ac{EM} system, as it will be shown in the next.
 Due to the presence of feedback, they impose the boundary conditions that are responsible for the presence of field discontinuity.  This aspect will be discussed in Sec. \ref{Sec:D}.

\subsection{System EM Transfer Function}
\label{Sec:STF}

It is of interest to evaluate the (space-variant) system \ac{EM} transfer function $\TildeGreenTot(\wavefreq,\wavefreqp)$ (or system Green function) that relates the impressed currents (input) and the resulting \ac{EM} field (output). 
Typically, only the impressed electric currents and the electric field are considered then, without loss of generality, we focus on the \ac{EM} transfer function component $\TildeGreenTotej(\wavefreq,\wavefreqp)$ that relates $\TildeJmi(\wavefreq)=\sum_m \TildeJmi^{(m)}(\wavefreq)$ and the electric field $\TildeEm(\wavefreq)$. In general, being the system space variant,  $\TildeEm(\wavefreq)$ can be expressed  as
\begin{align} \label{eq:Ek}
\TildeEm(\wavefreq)= \frac{1}{(2\pi )^3} \intthree \TildeGreenTotej(\wavefreq,\wavefreqbis)  \, \TildeJmi(\wavefreqbis) \, d \wavefreqbis  \, .
\end{align}

The system transfer function $\TildeGreenTotej(\wavefreq,\wavefreqp)$  indicates what is the response of the entire system at wavenumber $\wavefreq$ when it is solicited by an impressed current with wavenumber $\wavefreqp$. Specifically, the component $\TildeGreenTotejxx(\wavefreq,\wavefreqp)$  of dyadic $\TildeGreenTotej(\wavefreq,\wavefreqp)$ represents the response of the system at polarization $\versora_x$ when solicited by the \emph{harmonic current} $ \TildeJmi(\wavefreq) =\versora_x (2 \pi)^3 \delta(\wavefreq-\wavefreqp)$. Similarly for the other polarization combinations.
The  harmonic current is a dual of the infinitesimal source current, and it has only a mathematical meaning. 

It is customary to evaluate the electric field  observed along a plane, for instance, the $x-y$ plane at $z=z_{\text{o}}$ (not containing current sources), namely $\TildeEm(\wavefreqx,\wavefreqy; z_{\text{o}})$. For example, this could be the plane where a receiving antenna array or a surface is located. 
As a consequence, we can define the \ac{EM} system transfer function or channel transfer function \cite{PooBroTse:05,PizSanMar:22,PizSanMar:22a,PizLozRanMar:23}, namely $\Htf(\wavefreqx,\wavefreqy,\wavefreqxp,\wavefreqyp; z_{\text{s}}\, ,z_{\text{o}})$, with reference to a source in the $x-y$ plane at $z=z_{\text{s}}$, where the impressed (real or equivalent) currents ${\Jmi}(x,y; z_{\text{s}})$ are supposed to lay, and the  $x-y$ observation plane at $z=z_{\text{o}}$ \cite{PooBroTse:05,PizSanMar:22,PizSanMar:22a,PizLozRanMar:23}. 
It follows that 
\begin{align} \label{eq:Gtot}
\TildeEm(\wavefreqx,\wavefreqy; z_{\text{o}})&=
\frac{1}{2\pi}
\int  \TildeEm(\wavefreq) \, e^{\jmath \wavefreqz \,z_{\text{o}} } \, d \wavefreqz 
=
\inttwo \Htf(\wavefreqx,\wavefreqy,\wavefreqbisx,\wavefreqbisy; z_{\text{s}}\, ,z_{\text{o}})\, \TildeJmi(\wavefreqbisx,\wavefreqbisy;z_{\text{s}})  \, d \wavefreqbisx \, d \wavefreqbisy
\end{align}
where
\begin{align} \label{eq:Htf}
\Htf(\wavefreqx,\wavefreqy,\wavefreqxp,\wavefreqyp; z_{\text{s}}\, ,z_{\text{o}})&
=\frac{1}{(2 \pi)^2} \inttwo \TildeGreenTotej(\wavefreq,\wavefreqp) \, e^{-\jmath \wavefreqz \, z_{\text{s}} } e^{\jmath \wavefreqzp \, z_{\text{o}} }  d\wavefreqz \, \wavefreqzp\, .
\end{align} 
Analogously to
$\TildeGreenTotej(\wavefreq,\wavefreqp)$, the component $\Htfxx(\wavefreqx,\wavefreqy , 
\wavefreqxp,\wavefreqyp; z_{\text{s}}\, ,z_{\text{o}})$ of \eqref{eq:Htf} gives the system 
response observed on the plane $z=z_{\text{o}}$  at the 2D wavenumber $(\wavefreqx,\wavefreqy)$ and polarization $\versora_x$ when solicited by the harmonic current 
$\TildeJmi(\wavefreq) =\versora_x (2 \pi)^2 \delta(\wavefreqx-\wavefreqxp) \, \delta(\wavefreqy-\wavefreqyp) \, e^{-\jmath \wavefreqz z_{\text{s}}}$ located on the plane  $z=z_{\text{s}}$. 
 Note that \eqref{eq:Htf} is a vectorial transfer function, whereas the treatment in \cite{PooBroTse:05,PizSanMar:22,PizSanMar:22a} consider scalar fields.

In the following, we present an approach, based on linear algebra, to compute efficiently the \ac{EM} transfer function $\TildeGreenTotej(\wavefreq,\wavefreqp)$ in \eqref{eq:Ek}
or $\Htf(\wavefreqx,\wavefreqy,\wavefreqxp,\wavefreqyp; z_{\text{s}}\, ,z_{\text{o}})$ in \eqref{eq:Htf},  
 which allows to easily incorporate 
design, analysis, and optimization problems involving reconfigurable \acp{EMO}.
The approach permits also the evaluation of the \ac{EM} field as well as the impressed currents on each \ac{EMO}'s surface.

\section{Linear Algebra Formulation}
\label{Sec:Linear}


The following approach takes inspiration from the well-known method of moments  or mode matching \cite{Har:67}. 
For convenience, we introduce the inner product between vector functions $\Am(\boldr)$ and $\Bm(\boldr)$, defined on the generic surface $\Scal$, as 
\begin{align}
  \innerprod{ \Am (\boldr) }{ \Bm(\boldr)}= \int_{\Scal} \Am (\boldr) \scalprod \Bm^* (\boldr) \, d \boldr  \, .
\end{align}
Note that his definition is different from that used in the method of moments \cite{Rum:54}.
 Suppose  $\left \{ \Phim^{(m)}_n (\boldr) \right \}_{n=1,2, \ldots,   \Nm}$ is a complete vector orthonormal basis set for $\Scal^{(m)}$.
The orthogonality condition implies that
\begin{align} \label{eq:orth}
 \innerprod{ \Phim^{(m)}_n (\boldr)}{ \Phim^{(m)}_i (\boldr) }= \delta_{n,i} \, .
\end{align}
It is worth noticing  that $\Phim^{(m)}_n (\boldr)$ is a 3D vector which is tangent to the surface $\Scal^{(m)}$ for all $\boldr \in \Scal^{(m)}$ and zero otherwise.
Moreover, all the basis sets refer to disjoint surfaces so that $\left < \Phim^{(m)}_u (\boldr) , \Phim^{(i)}_n (\boldr) \right >=0$, $\forall u,n$ and $i \ne m$.
It follows that any vector function (field) $\Am(\boldr)$ lying on surface $\Scal^{(m)}$ can be represented as a linear combination of the basis functions (\emph{modes}) composing the basis set\footnote{In general, $\Nm$ could be infinity for the basis set to be complete. In such a case, $\Nm$ can be set to a finite value sufficiently large according to the desired level of accuracy. } 
\begin{align}
\Am(\boldr) & = \sum_{n=1}^{\Nm} a_n \, \Phim_n(\boldr) 
\end{align}
where the complex coefficients $\{ a_n\}$ are given by
\begin{align}
  a_n= \left < \Am(\boldr) , \Phim^{(m)}_n (\boldr) \right >  \, \, \, \, \, \, \, \, \, \, \, \quad n=1,2, \ldots, \Nm \, .
  \end{align}
Accordingly, the components $\Jm^{(m)}(\boldr)$  and $\Mm^{(m)}(\boldr)$ in \eqref{eq:JmMn}  can be represented in terms of the series expansions
\begin{align}
\label{eq:Jmm}
\Jm^{(m)}(\boldr) & = \sum_{n=1}^{\Nm} \bjmn \, \Phim_n^{(m)}(\boldr) + \sum_{n=1}^{\Nm} \ajmn \, \Phim_n^{(m)}(\boldr)  \\
 \Mm^{(m)}(\boldr) & = \sum_{n=1}^{\Nm} \bmmn \, \Phim_n^{(m)}(\boldr) + \sum_{n=1}^{\Nm} \ammn \, \Phim_n^{(m)}(\boldr)  
 \label{eq:Mmm}
\end{align}
where $\ajmn=\innerprod{  \Jmi^{(m)}(\boldr)}{ \Phim^{(m)}_n (\boldr) }$, $\bjmn=\innerprod{  \Jms^{(m)}(\boldr)}{ \Phim^{(m)}_n (\boldr)}$, 
$\ammn=\innerprod{  \Mmi^{(m)}(\boldr)}{ \Phim^{(m)}_n (\boldr) }$, and $\bmmn=\innerprod{  \Mms^{(m)}(\boldr)}{ \Phim^{(m)}_n (\boldr)}$.
Denote with $\ajm=\left [ \left \{ \ajmn \right \} \right ]$,  $\bjm=\left [ \left \{ \bjmn \right \} \right ]$, $\amm=\left [ \left \{ \ammn \right \} \right ]$, $\bmm=\left [\left \{ \bmmn \right \} \right ]$ the column vectors collecting the coefficients in \eqref{eq:Jmm} and \eqref{eq:Mmm}, respectively. We define also the vectors $\amv=\left [  \ {\ajm}^T \, \, {\amm}^T \right ]^T$ and $\bmv=\left [ {\bjm}^T \,  \, {\bmm}^T \right ]^T$.

By applying the inner product to both sides of 
\eqref{eq:EmHm} with the $n$th basis function $\Phim^{(m)}_n(\boldr)$ of the generic $m$th \ac{EMO}, and  by exploiting \eqref{eq:Jmm}-\eqref{eq:Mmm} as well as the orthogonality condition \eqref{eq:orth}, we obtain
\begin{align} 
e_n^{(m)}=& \innerprod{  \Em(\boldr)}{ \Phim^{(m)}_n (\boldr) }  =\innerprod{  \Emmt (\boldr)}{ \Phim^{(m)}_n (\boldr) }=
\innerprod{ \Greenej \cdot \Jm(\boldr)}{\Phim^{(m)}_n (\boldr)} + \innerprod{ \Greenem \cdot   \Mm(\boldr)}{\Phim^{(m)}_n (\boldr)}  \nonumber   \\
= &  \sum_{i=1}^M \innerprod{ \Greenej \cdot \Jm^{(i)}(\boldr)}{\Phim^{(m)}_n (\boldr)} + \sum_{i=1}^M \innerprod{ \Greenem \cdot  \Mm^{(i)}(\boldr)}{\Phim^{(m)}_n (\boldr)} \nonumber \\
= &  \sum_{i=1}^M \sum_{u=1}^{\Ni} \left (\ajiu + \bjiu \right ) \innerprod{ \Greenej  \cdot \Phim^{(i)}_u (\boldr)}{\Phim^{(m)}_n (\boldr)} + \sum_{i=1}^M \sum_{u=1}^{\Ni} \left ( \amiu + \bmiu \right ) \innerprod{ \Greenem \cdot  \Phim^{(i)}_u (\boldr)}{\Phim^{(m)}_n (\boldr)} 
\label{eq:enm}
\end{align}  
Similarly for the magnetic field component
\begin{align} \label{eq:hnm}
 h_n^{(m)} =& \innerprod{  \Hm(\boldr)}{ \Phim^{(m)}_n (\boldr) }  =\innerprod{  \Hmmt (\boldr)}{ \Phim^{(m)}_n (\boldr) } \nonumber \\
 = &  \sum_{i=1}^M \sum_{u=1}^{\Ni} \left ( \ajiu + \bjiu \right ) \innerprod{ \Greenhj \cdot \Phim^{(i)}_u (\boldr)}{\Phim^{(m)}_n (\boldr)} + \sum_{i=1}^M \sum_{u=1}^{\Ni}\left (\amiu + \bmiu \right ) \innerprod{ \Greenhm \cdot \Phim^{(i)}_u (\boldr)}{\Phim^{(m)}_n (\boldr)}   . 
\end{align}
As a consequence, the \ac{EM} field  $\left (\Emmt (\boldr),\Hmmt (\boldr) \right )$ tangent to the surface $\Scal^{(m)}$ can be expressed according to the series expansions
\begin{align}
\Emmt (\boldr) & = \sum_{n=1}^{\Nm} e_n^{(m)} \, \Phim^{(m)}_n(\boldr) 
&
\Hmmt (\boldr) & = \sum_{n=1}^{\Nm} h_n^{(m)} \, \Phim^{(m)}_n(\boldr) 
\label{eq:EHmt} \, .
\end{align}
Note that the above series expansion is valid only for the \ac{EM} tangential to the surface. 
Define the vector $\fmv=\left [ {\emv}^T \, {\hmv}^T  \right ]^T$ of dimension $2 \Nm$, with $\emv=\left [ \left \{  e_n^{(m)} \right \} \right ]$ and $\hmv=\left [ \left \{  h_n^{(m)} \right \} \right ]$ column vectors collecting the coefficients in \eqref{eq:EHmt}. By considering \eqref{eq:enm} and \eqref{eq:hnm},  $\fmv$ can be written in matrix form as
\begin{align} \label{eq:fmv}
	\fmv&=\sum_{i=1}^M \boldG^{(m,i)} \left [ \biv + \aiv \right ]
\end{align}
where
 \begin{equation}
\boldG^{(m,i)}= 
\left [
\begin{array}{cc}
  \boldG^{(m,i)}_{\text{EJ}} & \boldG^{(m,i)}_{\text{EM}}  \\
   \boldG^{(m,i)}_{\text{HJ}} & \boldG^{(m,i)}_{\text{HM}} \\  
     \end{array} 
\right ]
\end{equation}
is the coupling matrix of dimension $2 N^{(m)} \times 2 N^{(i)}$, whose elements are given by 
\begin{align} \label{eq:g11}
	 \left [ \boldG^{(m,i)}_{\text{EJ}} \right ]_{u,n}&=
	 \innerprod{\Greenej \Phim^{(i)}_u (\boldr)  }{\Phim^{(m)}_n (\boldr)} \\
	 &= \frac{1}{\jmath \omega \epsilon} \int_{\Scal^{(m)}}   \left (  \Phim^{(m)}_n (\boldr) \right )^* \scalprod \curl \curl \int_{\Scal^{(i)}} \Green (\boldr - \bolds) \, \Phim^{(i)}_u (\bolds) \, d\bolds \, d\boldr \\
	 \left [ \boldG^{(m,i)}_{\text{EM}} \right ]_{u,n}&=
	 \innerprod{\Greenem  \Phim^{(i)}_u (\boldr)  }{\Phim^{(m)}_n (\boldr)} \\
	 &= -  \int_{\Scal^{(m)}}   \left (  \Phim^{(m)}_n (\boldr) \right )^* \scalprod  \curl \int_{\Scal^{(i)}} \Green (\boldr - \bolds) \, \Phim^{(i)}_u (\bolds) \, d\bolds \, d\boldr \\
 \left [ \boldG^{(m,i)}_{\text{HJ}} \right ]_{u,n}&=
	 \innerprod{\Greenhj \Phim^{(i)}_u (\boldr)  }{\Phim^{(m)}_n (\boldr)} =-\left [ \boldG^{(m,i)}_{\text{EM}} \right ]_{u,n} \\
\left [ \boldG^{(m,i)}_{\text{HM}} \right ]_{u,n}&=
	 \innerprod{\Greenhm \Phim^{(i)}_u (\boldr)  }{\Phim^{(m)}_n (\boldr)} =\frac{\epsilon}{\mu}\left [ \boldG^{(m,i)}_{\text{EJ}} \right ]_{u,n}
	 \label{eq:g22}	 	 
	\end{align}
for $n=1,2, \ldots ,\Nm \,  , \, u=1,2,  \ldots,  \Ni$.  
It is worth noticing that the coupling matrices above depend only on the reciprocal geometry between \acp{EMO}  $i$ and $m$, i.e., their relative position and orientation. When $i=m$ (self-coupling), they depend neither on the position nor on the orientation.  
When the coupling between $m$ and $i$ is negligible, it is $\boldG^{(m,i)}_{\text{EJ}}, \boldG^{(m,i)}_{\text{EM}}, \boldG^{(m,i)}_{\text{HM}}, \boldG^{(m,i)}_{\text{HJ}} \approx \mathbf{0}_{2 N^{(m)} \times 2 N^{(i)}}$.
Following a similar approach,  also the constitutive equation \eqref{eq:JMs} can be put in matrix form
\begin{equation} \label{eq:bmv}
	\bmv=\boldD^{(m)} \, \fmv
\end{equation}
where
\begin{equation} \label{eq:Dm}
\boldD^{(m)}= 
\left [
\begin{array}{cc}
  \boldD^{(m)}_{\text{JE}} & \boldD^{(m)}_{\text{JH}}  \\
   \boldD^{(m)}_{\text{ME}} & \boldD^{(m)}_{\text{MH}} \\  
     \end{array} 
\right ]
\end{equation}
and
\begin{align}
\left [  \boldD^{(m)}_{\text{JE}} \right ]_{u,n}&=\innerprod{\Dmje \cdot \Phim^{(m)}_u (\boldr)  }{\Phim^{(m)}_n (\boldr)} \\
\left [  \boldD^{(m)}_{\text{JH}} \right ]_{u,n}&=\innerprod{\Dmjh \cdot \Phim^{(m)}_u (\boldr)  }{\Phim^{(m)}_n (\boldr)} \\
\left [  \boldD^{(m)}_{\text{ME}} \right ]_{u,n}&=\innerprod{\Dmme \cdot \Phim^{(m)}_u (\boldr)  }{\Phim^{(m)}_n (\boldr)} \\
\left [  \boldD^{(m)}_{\text{MH}} \right ]_{u,n}&=\innerprod{\Dmmh \cdot \Phim^{(m)}_u (\boldr)  }{\Phim^{(m)}_n (\boldr)} 
\end{align} 
 with $u,n=1,2, \ldots \Nm$.
Matrix $\boldD^{(m)}$, of dimension $2 N^{(m)} \times 2 N^{(m)}$,  describes completely the linear transformation operated by the $m$th \ac{EMO}, polarization effects included, under the limit of the series expansion approximation. 

For instance, if we are interested in finding the \ac{EM} tangential on surface $\Scal^{(m)}$ of the $m$th \ac{EMO},  by combining \eqref{eq:fmv} and \eqref{eq:bmv},  it is
 \begin{align} \label{eq:fmv1}
	\fmv&=\sum_{i=1}^M \boldG^{(m,i)}  \boldD^{(i)} \fiv +   \sum_{i=1}^M \boldG^{(m,i)}  \aiv  \\
	\fmv&= \left ( \boldI - \boldG^{(m,m)}  \boldD^{(m)} \right )^{-1}  \, 
	\left (  \sum_{i=1, i\ne m}^M \boldG^{(m,i)}  \boldD^{(i)} \fiv +   \sum_{i=1}^M \boldG^{(m,i)}  \aiv \right )  \, .
\end{align}

\begin{figure}[t]
\centerline{\includegraphics[width=0.8\columnwidth]{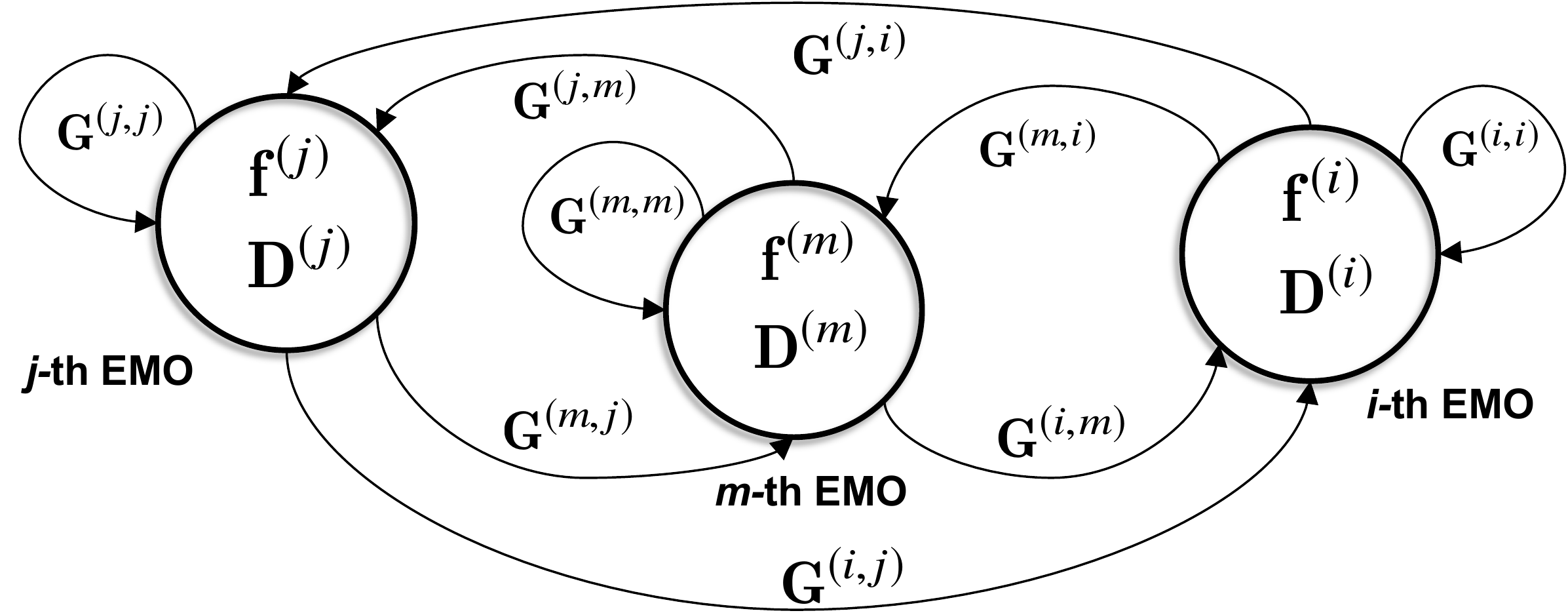}}
\caption{\ac{EM} scenario as a fully connected graph.}
\label{Fig:EMGraph}
\end{figure}

The matrix relationships above can be graphically represented as a connected graph sketched  in Fig. \ref{Fig:EMGraph}. Graph theory tools, such as the Mason's gain formula \cite{Mason:56}, can be exploited to solve \eqref{eq:fmv1} or any other set of equations  depending on the structure of the system. 
Significant simplifications can be operated if the coupling between some \acp{EMO} is weak and/or the induced currents at any \ac{EMO} are negligible. 

Another example is the derivation of the relationship between the coefficients of the  incident \ac{EM} field $\fmv_{\text{inc}}$ and those of the  scattered field $\fmv_{\text{s}}$ at the generic $m$th \ac{EMO}. In particular, it is
\begin{equation}
\fmv_{\text{s}}=\left (\boldI- \boldG^{(m,m)} \boldD^{(m)} \right )^{-1} \boldG^{(m,m)} \boldD^{(m)}  \, \fmv_{\text{inc}}
\end{equation}
where $\fmv_{\text{s}}+\fmv_{\text{inc}}=\fmv$ and $\fmv_{\text{inc}}=\sum_{i=1, i\ne m}^M \boldG^{(m,i)}  \boldD^{(i)} \fiv +   \sum_{i=1}^M \boldG^{(m,i)}  \aiv$.
Matrix representation is useful in optimization problems where the best configuration of one or more \acp{EMO}, i.e., their matrix $\boldD$, must be found to achieve a given result on the \ac{EM} field  under some constraints such as the total power. 
Some examples will be provided in Sec. \ref{Sec:Examples}.

\section{Basis Sets and Sources}
\label{Sec:Basis}

A crucial aspect of the linear algebra formulation illustrated in the previous section is the choice of the basis functions that affect the trade-off between accuracy and computational complexity, i.e., their number \cite{BalB:16,ChenB:18}. 
We choose the harmonic basis functions,  which allow for an efficient representation and the possibility to exploit the properties of the harmonic analysis.
Without loss of generality, we consider \acp{EMO} oriented according to the plane $z=0$ and centered at the origin (canonical position and orientation). How to deal with differently oriented and positioned \acp{EMO} will be explained at the end of the section. In the following, we illustrate possible basis sets for some geometries of interest.

\subsection{Infinitesimal Vertically Polarized Current Source}
\label{Sec:pointsource}
Considering a canonical vertical polarization, the only possible base function  is $\Phim_1(\boldr)=\versory \, \delta(x) \, \delta(y) \, \delta(z)$ so that $\TildePhim_1(\wavefreq)=\fourier{\Phim_1(\boldr)}=\versory$. 
The infinitesimal source is typically used to model the small dipole of infinitesimal length $\Delta L$ and current $I_0$ whose current density can be written as $\Jm(\boldr)=\versory I_0 \, \Delta L  \, \Phim_1(\boldr)$  and $\TildeJm(\wavefreq)= I_0 \, \Delta L  \, \TildePhim_1(\wavefreq)$ (Hertzian dipole) \cite{BalB:16}. 

\subsection{Line Source of Length $L$}

\begin{figure}[t]
\centerline{\includegraphics[width=0.65\columnwidth]{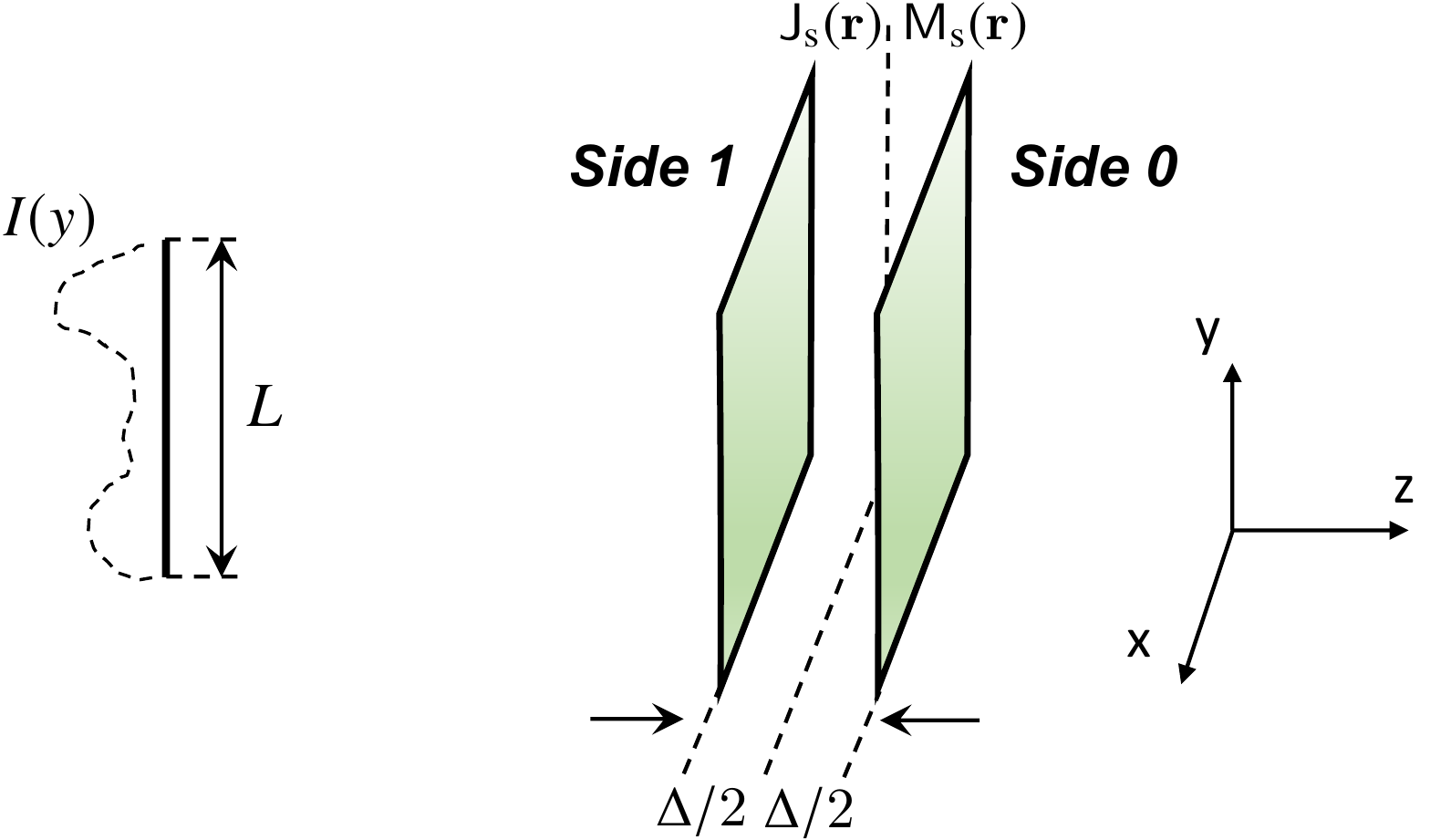}}
\caption{Line source (left) and thin surface model (right). }
\label{Fig:SurfaceModel}
\end{figure}

An example of line source, shown in Fig. \ref{Fig:SurfaceModel} left, is given by the conducting wire of length $L$. In this case we have $\Jm(\boldr)=\versory \, I(y)\,  \delta(x) \, \delta(z)$,  with current distribution $I(y)$ different from zero in $|y|<L/2$. 
The basis functions  are $\Phim_n(\boldr)=\versory \, \phi_{n}(\boldr)$, where $\phi_{n}(\boldr)$ are scalar basis functions for  $n=1,2, \ldots, N_y$ ($N_y$ odd number).
For convenience and with some abuse of notation, consider also the following alternative indexing 
$\Phim_{n_y}(\boldr)$, where $n_y$ is related to $n$ according to the mapping $n_y=n-(N_y-1)/2-1$, for $n_y=-(N_y-1)/2, \ldots, -1, 0, 1, \ldots, (N_y-1)/2$.
A complete basis set for a vertical line of length $L$, with    $y \in [L/2,L/2]$, is given by $\phi_{n_y}(\boldr)=I_{n_y}(y;L)  \,  \delta(x) \, \delta(z)$, having defined $I_{n_y}(y;L)=\frac{1}{\sqrt{L}} \rect{\frac{y}{L} } \exp{ \left ( \jmath  \frac{2 \pi n_y y}{L} \right )}$, where  the coefficient ensures that the basis functions have  unitary energy. In the frequency domain, we have  $\TildePhim_{n_y}(\wavefreq)=\versory \, \tilde{\phi}_{n_y} (\wavefreq;L)$, with $\tilde{\phi}_{n_y} (\wavefreq; L)=\tilde{\phi}_{n_y} (k_y ; L)=S_{n_y}(k_y ; L) $, where 
\begin{equation}  \label{eq:Sn}
S_n(k; L) =\sqrt{L} \, \sinc{\frac{k \, L}{2 \pi}-n  }   \, .
\end{equation}

\subsection{Thin Surface of Size $L_x \times L_y$}
\label{Sec:surfacemodel}


A thin \ac{RIS} that is electrically large and made of sub-wavelength reconfigurable scattering elements, is homogenizable and can be modeled as a continuous surface sheet characterized by suitable surface functions (in general dyadic tensors) such as impedances or admittances  (impedance sheet) \cite{DiRDanTre:22}.
This allows to express  the boundary conditions introduced by the surface at a macroscopic level,  thus abstracting the microscopic structure of the surface and leading to a good  compromise between accuracy and model tractability. 
Therefore, it is of interest to define a basis set to express any surface function. 
By applying the basis functions  in \eqref{eq:Sn}  to each dimension of the surface, any scalar surface function can be represented as a linear combination of the following scalar basis functions
 \begin{align} 
 \phi_{n_x,n_y}(\boldr)&=I_{n_x}(x; L_x)  \, I_{n_y} (y; L_y) \, \delta(z) 
 &
 \tilde{\phi}_{n_x,n_y}(\wavefreq)&=S_{n_x}(k_x; L_x) \, S_{n_y}(k_y; L_y)
 \end{align}
respectively, in the spatial and wavenumber domains, 
with $n_x=-(N_x-1)/2, \ldots, -1, 0, 1, \ldots, (N_x-1)/2$, $n_y=-(N_y-1)/2, \ldots, -1, 0, 1, \ldots, (N_y-1)/2$.

The currents induced by the incident \ac{EM} field introduce a discontinuity in the \ac{EM} fields   
on the two sides of the surface. Therefore, we model it with two separate faces (sides) at infinitesimal distance $\Delta$ at $z=-\Delta/2$ and $z=\Delta/2$, where $\Delta\ll \lambda$ is the thickness of the surface, as shown in Fig. \ref{Fig:SurfaceModel} right.  
To condense the notation, with introduce the alternative indexing $n=n_x+ (N_x-1)/2)+1 + N_x (n_y + (N_y-1)/2) )+ \np N_x N_y + \ns 2 N_x N_y$, where  $\np=0$ if horizontally polarized and $\np=1$ if vertically polarized, $\ns=0$ if  right-side face, and $\ns=1$ if   left-side face.  In this manner, the complete basis set necessary to represent any \ac{EM} field  on the two sides of the surface is 
\begin{align}
\Phim_n(\boldr)=\Phim_{n_x,n_y,\np,\ns}(\boldr)&=\versora_{\np} \, \phi_{n_x,n_y}(\boldr+(0.5-\ns) \, \Delta\,  \versorz) \\
\TildePhim_{n_x,n_y,\np,\ns}(\wavefreq)&=\versora_{\np} \, \tilde{\phi}_{n_x,n_y} (\wavefreq) \, e^{-\jmath \, \wavefreqz\,  (0.5-\ns)\, \Delta}
\end{align}
where $\versora_{\np}=\versorx$, when $\np=0$, and $\versora_{\np}=\versory$, when $\np=1$.
Therefore, the total number of basis functions is  $N=4 N_x N_y$. 
Currents are supposed to lay at $z=0$, i.e., in the middle of the two sides,  to avoid singularities, and hence $2 N_x N_y$  basis functions are sufficient to represent them.

\subsection{Basis Functions for Plane Waves and Elementary Harmonic Currents}
 
 In order to create a  ``bridge" between the linear algebra  characterization in Sec. \ref{Sec:Linear} and the transfer function characterization in \eqref{eq:Ek}  and \eqref{eq:Htf} of the \ac{EM} system, it is convenient to define a virtual \ac{EMO} consisting of a generic plane wave with wavenumber $\wavefreqp=(\wavefreqxp,\wavefreqyp,\wavefreqzp)$  and polarization $\versora(\wavefreqp)$. As it will be clearer later, thanks to this virtual \ac{EMO}, it is possible to determine how the \acp{EMO} 
  are coupled with the \ac{EM} field at the generic wavenumber $\wavefreqp$.
  In other words, it can be used to ``observe" the \ac{EM} field without any influence on it. Any plane wave can be fully represented as a linear combination of two basis functions $\Phim_n(\boldr; \wavefreqp)=\versora_n(\wavefreqp) \, e^{\jmath \wavefreqp \scalprod \boldr}$, 
 where the vectors $\versora_1(\wavefreqp)$ and $\versora_2(\wavefreqp)$  are, respectively, transversal  and longitudinal with respect to the direction of propagation $\wavefreqp$ since it should be $\wavefreqp \scalprod \versora(\wavefreqp)=0$ \cite{CleB:96}. 
 Taking the Fourier transform, it is 
 \begin{equation} \label{eq:Plane}
  \TildePhim_n(\wavefreq ; \wavefreqp)=(2\pi)^3 \versora_n(\wavefreqp)
 \, \delta(\wavefreq-\wavefreqp)  \, .
 \end{equation}
 
 
It is also of utility the definition of the plane wave observed on the $x-y$ plane at $z=z_{\text{o}}$, that is, $\Phim_n(\boldr; \wavefreqp, z_{\text{o}})=\versora_n(\wavefreqp) \, \delta(z-z_{\text{o}}) \, e^{\jmath \wavefreqp \scalprod \boldr}$, with $\versora_1(\wavefreqp)=\versorx$ and $\versora_2(\wavefreqp)=\versory$, whose 2D Fourier transform,  for $n=1,2$, 
 is
 \begin{equation} \label{eq:Planez}
  \TildePhim_n(\wavefreqx, \wavefreqy ; \wavefreqp, z_{\text{o}} )=(2\pi)^2 \versora_n(\wavefreqp)
 \, \delta(\wavefreqx-\wavefreqxp) \, \delta(\wavefreqy-\wavefreqyp) \, e^{-\jmath \wavefreqzp \, z_{\text{o}}} \, .
 \end{equation}
 Analogously, we can define the elementary harmonic electric current with polarization $\versora$ and wavenumber $\wavefreqp$ flowing on the $x-y$ plane at $z=z_{\text{s}}$ having the 2D Fourier transform
\begin{equation} \label{eq:JPlanez}
  \TildePhim(\wavefreqx, \wavefreqy ; \wavefreqp, z_{\text{s}} )=(2\pi)^2 \versora
 \, \delta(\wavefreqx-\wavefreqxp) \, \delta(\wavefreqy-\wavefreqyp) \, e^{-\jmath \wavefreqzp \, z_{\text{s}}} \, .
 \end{equation}

\subsection{Power Flux through a Surface}

In optimization problems, often it is important to include constraints on power budget. For instance, in a passive \ac{RIS} the reflected radiated power cannot be larger than the power of the incident field. The complex power flux flowing through a surface is given by  
\begin{align}
	P&=\frac{1}{2}  \int_{\Scal} \Emmt (\boldr) \times \left (\Hmmt (\boldr) \right )^*   \, d\boldr =\frac{1}{2} 
	\sum_{n=1}^{\Nm} \sum_{i=1}^{\Nm} e_n^{(m)} \left (h_i^{(m)} \right )^*
	 \int_{\Scal} \Phim^{(m)}_n(\boldr)  \times \Phim^{(m)}_i(\boldr)    \, d\boldr \nonumber \\
	 &=\frac{1}{2}  \sum_{n_x,n_y} \sum_{i_x,i_y} 
	 \left [e_{n_x,n_y,\np=0}^{(m)} \, \left (h_{i_x,i_y,i_{\text{p}}=1}^{(m)}\right )^*- e_{n_x,n_y,\np=1}^{(m)} \, \left (h_{i_x,i_y,i_{\text{p}}=0}^{(m)}\right )^* \right ] 
	\int_{\Scal} \phi_{n_x,n_y}(\boldr)  \, \phi_{i_x,i_y}(\boldr)    \, d\boldr  \nonumber \\
	&=\frac{1}{2} \sum_{n_x,n_y}
	 \left [e_{n_x,n_y,\np=0}^{(m)} \, \left (h_{n_x,n_y,\np=1}^{(m)}\right )^*- e_{n_x,n_y,\np=1}^{(m)} \, \left (h_{n_x,n_y,\np=0}^{(m)}\right )^* \right ] \, .
	\end{align}

The real part of $P$, $P_{\text{rad}}=\Re \left \{P \right \} $, represents the radiated power, whereas the imaginary part the reactive power component.

\subsection{\acp{EMO} in Non-canonical Position and Orientation}

The  basis functions in the wavenumber domain for \acp{EMO} with different position $\boldp$ and orientation $\DR$ can be easily obtained by exploiting the classical Fourier property
\begin{equation} 
\TildePhim(\wavefreq)= \DR \cdot \OPhim \left (\DR \cdot \wavefreq \right ) e^{-\jmath \boldp \scalprod \wavefreq} 
\end{equation}
where $ \OPhim (\wavefreq)$ is the basis function in the wavenumber domain of the \ac{EMO} in the canonical position and orientation. The multiplication by the dyadic $\DR$ ensures a correct  polarization rotation. For instance, the rotation matrix corresponding to a generic rotation angle $\theta$ around the $y$-axis is given by 
\begin{equation}
\DR_y(\theta)=\left [
\begin{array}{ccc}
 \cos \theta & 0  & \sin \theta   \\
 0 &  1 &  0 \\
 -\sin \theta & 0  &   \cos \theta
\end{array}
\right ] \, .
\end{equation}

\section{Modes Coupling}
\label{Sec:Coupling}

The last ingredient necessary to implement the method based on linear algebra is the derivation of the coupling coefficients between modes composing the matrices $\boldG^{(m,i)}$. 
Since the coupling between basis functions belonging to the $i$th and $m$ \acp{EMO} depends only on their relative position and orientation, it is convenient to consider $\TildePhim_u^{(i)}(\wavefreq)=\OPhim_u^{(i)}(\wavefreq)$ located at the origin  
 (canonical position and orientation), and  $\TildePhim_n^{(m)}(\wavefreq)$ at the relative position and orientation $\boldp=\boldp^{(m,i)}=\boldp^{(m)}-\boldp^{(i)}$ and $\DR^{(m,i)}$, respectively,  such that the two \acp{EMO} do not intersect along the $z$ axis. 
By applying the Parseval's theorem to \eqref{eq:g11}-\eqref{eq:g22}, it is
\begin{align} 
	 \left [ \boldG^{(m,i)}_{\text{EJ}} \right ]_{u,n}&=\frac{\jmath }{ \omega \epsilon (2 \pi)^3} \intthree \TildeGreen(\wavefreq) \, \left ( \TildePhim_n^{(m)}(\wavefreq) \right )^*   \scalprod  \left ( \crossprod{\wavefreq}{\crossprod{\wavefreq}{  \TildePhim^{(i)}_u(\wavefreq)  }} \right )
   \, d \wavefreq  \nonumber \\ \label{eq:Gejk}
   &= \frac{\jmath }{ \omega \epsilon (2 \pi)^3} \intthree \TildeGreen(\wavefreq) \,  e^{\jmath \boldp^{(m,i)} \scalprod \wavefreq} \, \left ( \DR^{(m,i)} \cdot \OPhim_n^{(m)}\left (\DR^{(m,i)} \cdot \wavefreq \right ) \right )^*   \scalprod  \left ( \crossprod{\wavefreq}{\crossprod{\wavefreq}{  \OPhim^{(i)}_u(\wavefreq)  }} \right )
   \, d \wavefreq 
   \\ 
    \left [ \boldG^{(m,i)}_{\text{EM}} \right ]_{u,n}&=- \frac{\jmath }{(2 \pi)^3} \intthree \TildeGreen(\wavefreq) \, \left ( \TildePhim_n^{(m)  }(\wavefreq) \right )^*   \scalprod  \left ( \crossprod{\wavefreq}{   \TildePhim^{(i)}_u(\wavefreq)  } \right )
   \, d \wavefreq \nonumber \\
   &=- \frac{\jmath }{(2 \pi)^3} \intthree \TildeGreen(\wavefreq) \, e^{\jmath \boldp^{(m,i)} \scalprod \wavefreq} \left ( \DR^{(m,i)} \cdot \OPhim_n^{(m)}\left (\DR^{(m,i)} \cdot \wavefreq \right ) \right )^*   \scalprod  \left ( \crossprod{\wavefreq}{  \OPhim^{(i)}_u(\wavefreq)  } \right )
   \, d \wavefreq \, . \label{eq:Gemk} 
\end{align}

Using the equality \eqref{eq:Cauchy} in the Appendix, alternative expressions for \eqref{eq:Gejk} and \eqref{eq:Gemk} can be obtained, respectively, 
\begin{align} \label{eq:Gej2}
	 \left [ \boldG^{(m,i)}_{\text{EJ}} \right ]_{u,n}&=\frac{\pi}{\omega \epsilon (2 \pi)^3} \inttwo  \frac{e^{\jmath \boldp^{(m,i)} \scalprod \wavefreq^{\pm}}}{k_z(\wavefreqx,\wavefreqy)} \left ( \DR^{(m,i)} \cdot \OPhim_n^{(m)}\left (\DR^{(m,i)} \cdot \wavefreq^{\pm} \right ) \right )^*   \scalprod  \left ( \crossprod{\wavefreq^{\pm}}{\crossprod{\wavefreq^{\pm}}{  \OPhim^{(i)}_u(\wavefreq^{\pm})  }} \right )
   \, d \wavefreqx \, d \wavefreqy \\ \label{eq:Gem2}
	 \left [ \boldG^{(m,i)}_{\text{EM}} \right ]_{u,n}&=-\frac{\pi}{ (2 \pi)^3} \inttwo \frac{e^{\jmath \boldp^{(m,i)} \scalprod \wavefreq^{\pm}}}{k_z(\wavefreqx,\wavefreqy)} \left ( \DR^{(m,i)} \cdot \OPhim_n^{(m)}\left (\DR^{(m,i)} \cdot \wavefreq^{\pm} \right ) \right )^*   \scalprod  \left ( \crossprod{\wavefreq^{\pm}}{  \OPhim^{(i)}_u(\wavefreq^{\pm})  } \right )
   \, d \wavefreqx \, d \wavefreqy
   \end{align}
where $k_z=k_z \left (\wavefreqx, \wavefreqy \right )$ and $\wavefreq^{\pm}$ are defined in \eqref{eq:kz} and \eqref{eq:kpm}, respectively, with $r_z=p_z$ and $z_{\text{max}}=z_{\text{min}}=0$.
These alternate expressions are useful because they reduce the evaluation of the coupling coefficient to a 2D Fourier-type integral which can be numerically solved using efficient FFT tools  \cite{SheGloSanVar:89}.
In the following, we derive  further simplifications and closed-form expressions of the coupling coefficients for particular cases of interest.

\subsection{Coupling Between any \acp{EMO} in Far Field }

When two \acp{EMO} are located in their respective far-field region, i.e.,  $L^2/|\boldp| \ll \lambda$, where $L$ is the dimension of the largest \ac{EMO} and $\boldp=\boldp^{(m.i)}$, the approximation \eqref{eq:Approx} in the Appendix  can be applied to \eqref{eq:Gej2} and \eqref{eq:Gem2} thus obtaining  
%
\begin{align}
	 \left [ \boldG^{(m,i)}_{\text{EJ}} \right ]_{u,n}& \simeq 
	 \jmath \, \kappazero \, \eta   \frac{e^{\jmath \, \kappazero \, |\boldp|}}{ 4 \pi \, |\boldp|}  \left ( \DR^{(m,i)} \, \OPhim_n^{(m)} \left (\DR^{(m,i)} \, \wavefreq_{\boldp} \right )  \right )^*   \scalprod  \left ( \crossprod{\versorp^{\pm}}{\crossprod{\versorp^{\pm}}{  \OPhim^{(i)}_u(\wavefreq_{\boldp})  }} \right ) \\
\left [ \boldG^{(m,i)}_{\text{EM}} \right ]_{u,n}& \simeq 
	 - \jmath \, \kappazero   \frac{e^{\jmath \, \kappazero \, |\boldp|}}{ 4 \pi \, |\boldp|} \left ( \DR^{(m,i)} \, \OPhim_n^{(m)} \left (\DR^{(m,i)} \, \wavefreq_{\boldp} \right )  \right )^*   \scalprod  \left ( \crossprod{\versorp^{\pm}}{  \OPhim^{(i)}_u(\wavefreq_{\boldp})  } \right ) 	 
\end{align}
being $\wavefreq_{\boldp} = \kappazero\, \versorp^{\pm}$, where $\versorp^{\pm}=(p_x,p_y,|p_z|)/| \boldp |$. Since $|\versorp^{\pm} |=1$, it is evident that only plane waves with $\left ( \wavefreqx, \wavefreqy \right ) \in \CalP$ contribute to the propagation in far field of the \ac{EM} wave.

\subsection{Coupling with Plane Waves}

When the  $m$th \ac{EMO} is a plane wave with wavenumber $\wavefreqp$ and polarization $\versora_n(\wavefreqp)$, the coupling coefficients between the plane wave and the $i$th \ac{EMO} can be obtained in closed form by substituting \eqref{eq:Plane} in \eqref{eq:Gejk} and \eqref{eq:Gemk}. 
Of more  interest is the coupling with the wave plane in \eqref{eq:Planez} at the observation plane $z=z_{\text{o}}$  that can be easily obtained in closed form 
\begin{align} \label{eq:Gejkplane}
	 \left [ \boldG^{(m,i)}_{\text{EJ}} \right ]_{u,n}&=\left [ \boldG^{(m,i)}_{\text{EJ}} \right ]_{u,n}(\wavefreqp)= \frac{e^{\jmath \boldp^{(m,i)} \scalprod \wavefreqp^{\pm} } }{2 \omega \epsilon \, k_z(\wavefreqxp,\wavefreqyp)}   \versora_n(\wavefreqp^{\pm})  \scalprod  \left ( \crossprod{\wavefreqp^{\pm}}{\crossprod{\wavefreqp^{\pm}}{  \OPhim^{(i)}_u(\wavefreqp^{\pm})  }} \right )
   \\ \label{eq:Gemkplane}
    \left [ \boldG^{(m,i)}_{\text{EM}} \right ]_{u,n}&=\left [ \boldG^{(m,i)}_{\text{EM}} \right ]_{u,n}(\wavefreqp)= - \frac{  \, e^{\jmath \boldp^{(m,i)} \scalprod \wavefreqp^{\pm}   }}{2\, k_z(\wavefreqxp,\wavefreqyp) }	   \versora_n(\wavefreqp^{\pm} ) \scalprod  \left ( \crossprod{\wavefreqp^{\pm}}{   \TildePhim^{(i)}_u(\wavefreqp^{\pm})  } \right ) 
\end{align}
for $n=1,2$, and $u=1,2, \ldots, \Ni$, where $\boldp^{(m,i)}=\left (-p_x^{(i)},-p_y^{(i)},z_{\text{o}}-p_z^{(i)} \right )$, and $\wavefreqp^{\pm}$ is defined in \eqref{eq:kpm}, with $r_z=z_{\text{o}} - p_z^{(i)}$ and $z_{\text{max}}=z_{\text{min}}=0$. 
%
It is worth noticing that the coefficients in \eqref{eq:Gejkplane} and \eqref{eq:Gemkplane} are function of $\wavefreqp$.

\subsection{Coupling Between an EMO and an Harmonic Current}

Similarly, closed-form expressions for the coupling coefficients can be obtained when the generic $m$th \ac{EMO} is coupled with the harmonic current in \eqref{eq:JPlanez}, that is, 

\begin{align} \label{eq:Gejkplanecurr}
	 \left [ \boldG^{(m,i)}_{\text{EJ}} \right ]_{u,n}(\wavefreqp) &= \frac{e^{\jmath \boldp^{(m,i)} \scalprod \wavefreqp^{\pm}}}{2 \omega \epsilon \,  k_z(\wavefreqxp,\wavefreqyp)} \left ( \DR^{(m,i)} \cdot \OPhim_n^{(m)}\left (\DR^{(m,i)} \cdot \wavefreqp^{\pm} \right ) \right )^*   \scalprod  \left ( \crossprod{\wavefreqp^{\pm}}{\crossprod{\wavefreqp^{\pm}}{  \versora }} \right )
    \\ \label{eq:Gemkplanecurr}
	 \left [ \boldG^{(m,i)}_{\text{EM}} \right ]_{u,n} (\wavefreqp)&=- \frac{e^{\jmath \boldp^{(m,i)} \scalprod \wavefreqp^{\pm}}}{2\, k_z(\wavefreqxp,\wavefreqyp)} \left ( \DR^{(m,i)} \cdot \OPhim_n^{(m)}\left (\DR^{(m,i)} \cdot \wavefreqp^{\pm} \right ) \right )^*   \scalprod  \left ( \crossprod{\wavefreqp^{\pm}}{  \versora  } \right )
   \end{align}
for $u=1$, where $\boldp^{(m,i)}=\left (p_x^{(m)},p_y^{(m)},p_z^{(m)}-z_{\text{s}} \right )$, and $\wavefreqp^{\pm}$ is defined in \eqref{eq:kpm}, with $r_z=p_z^{(m)}-z_{\text{s}} $ and $z_{\text{max}}=z_{\text{min}}=0$. Also in this case the coefficients are function of $\wavefreqp$.

\subsection{Self-coupling in Surfaces}

In this case, $\DR^{\, (m,m)} =1$ and $\boldp=\boldp^{(m,m)}=(0,0,p_z)$, with $p_z=(1-2 \ns) \, \Delta/2$ depending whether side $\ns=0$ or $\ns=1$ is considered. The basis functions associated to currents are located at $z=0$. 
As a consequence, \eqref{eq:Gej2} and \eqref{eq:Gem2} read
\begin{align} \label{eq:gej3}
	 \left [ \boldG^{(m,m)}_{\text{EJ}} \right ]_{u,n}&=\frac{\pi}{\omega \epsilon (2 \pi)^3} \inttwo 
	 \frac{k_{u,n}(\wavefreqx,\wavefreqy) \, e^{\jmath k_z(\wavefreqx,\wavefreqy) \Delta/2}}{k_z(\wavefreqx,\wavefreqy)}
	  \left (  \Ophim_n^{(m)}\left ( \wavefreq \right ) \right )^*   \cdot  \left  ( \Ophim^{(m)}_u(\wavefreq)  \right )
   \, d \wavefreqx \, d \wavefreqy \\ \label{eq:gem3}
	 \left [ \boldG^{(m,m)}_{\text{EM}} \right ]_{u,n}&=-\frac{\pi}{(2 \pi)^3}  
	 \inttwo 
	 \frac{k_{u,n}^{\star}(\wavefreqx,\wavefreqy) \, e^{\jmath k_z(\wavefreqx,\wavefreqy) \Delta/2}}{k_z(\wavefreqx,\wavefreqy)}
	  \left (  \Ophim_n^{(m)}\left ( \wavefreq \right ) \right )^*   \cdot  \left  ( \Ophim^{(m)}_u(\wavefreq)  \right )
   \, d \wavefreqx \, d \wavefreqy 
\end{align}
where index $n$ refers to the $n$th current basis  function (at $z=0$), and  index $u$ refers to the $u$th basis  function of the \ac{EM} field on both sides of the surfaces at $p_z=(1-2 \ns) \, \Delta/2$ depending on $\ns$. In addition,  $k_{u,n}(\wavefreqx,\wavefreqy)=-\wavefreqy^2-k_z^2(\wavefreqx,\wavefreqy)$, $k_{u,n}^{\star}(\wavefreqx,\wavefreqy)=0$ when $\versora_n=\versora_u=\versorx$,  $k_{u,n}(\wavefreqx,\wavefreqy)=-\wavefreqx^2-k_z^2$, $k_{u,n}^{\star}(\wavefreqx,\wavefreqy)=0$ when $\versora_n=\versora_u=\versory$, $k_{u,n}=\wavefreqx\, \wavefreqy$, $k_{u,n}^{\star}(\wavefreqx,\wavefreqy)= k_z(\wavefreqx,\wavefreqy)\,  (\np -\up) (1-2\ns)$ when $\versora_n \ne \versora_u$, with $\versora_n \in \left \{\versorx,\versory  \right \}$ and $\versora_u\in \left \{\versorx,\versory  \right \}$ being the polarization of $\OPhim_n^{(m)}(\wavefreq)$ and $\OPhim_u^{(m)}(\wavefreq)$, respectively. When the polarizations are equal (i.e., $\np= \up$), $k_{u,n}^{\star}(\wavefreqx,\wavefreqy)=0$ and hence $\left [ \boldG^{(m,m)}_{\text{EM}} \right ]_{u,n}=0$. When $\np \ne \up$, $n_x=u_x$ and $n_u=u_y$ we have
\begin{align}
	 \left [ \boldG^{(m,m)}_{\text{EM}} \right ]_{u,n}&=-   \frac{\pi \, (\np-\up) (1-2\ns) \, \delta_{u_x-n_x, \, u_y-n_y} 
	 }{(2 \pi)^3} \inttwo e^{\jmath k_z (\wavefreqx,\wavefreqy)\Delta/2}
	  \left (  \Ophim_n^{(m)}\left ( \wavefreq \right ) \right )^*   \cdot  \left  (\Ophim^{(m)}_u(\wavefreq)  \right )
   \, d \wavefreqx \, d \wavefreqy  \, .
\end{align}
%

\subsection{Self-coupling in Large Surfaces}

When $L_x, L_y \gg \lambda$,  the unitary energy functions $S_n(k;L)$ in \eqref{eq:Sn} composing $\Ophim^{(m)}$  tend to zero very quickly around their maximum value compared to the speed of variations  of the other terms of the integrand,  therefore, the latter can be approximated as constant  and the integrals \eqref{eq:gej3} and \eqref{eq:gem3} solved, thus obtaining 
\begin{align} \label{eq:Gej3}
	 \left [ \boldG^{(m,m)}_{\text{EJ}} \right ]_{u,n}&\simeq \frac{\delta_{u_x-n_x, \, u_y-n_y} }{2\, \omega \epsilon } 	 \frac{k_{u,n}^{(n)}}{k_z^{(n)}} \, e^{\jmath k_z^{(n)} \Delta/2}  \noindent \\
	 \label{eq:Gem3}
	  \left [ \boldG^{(m,m)}_{\text{EM}} \right ]_{u,n}& \simeq - \frac{ (\np-\up) (1-2\ns )\, \delta_{u_x-n_x, \, u_y-n_y} }{2} e^{\jmath k_z^{(n)} \Delta/2} 
\end{align}
%
with $k_x^{(n)}=\frac{2 \pi n_x}{L_x}$, $k_y^{(n)}=\frac{2 \pi n_y}{L_y}$, $k_z^{(n)}=k_z \left ( k_x^{(n)}, k_y^{(n)} \right )$, and $k_{u,n}^{(n)}=k_{u,n} \left ( k_x^{(n)}, k_y^{(n)} \right )$. 
When $L_x, L_y \rightarrow \infty$ \eqref{eq:Gej3} converges to $-\delta_{u_x-n_x, \, u_y-n_y} \, \eta \, 
	 e^{\jmath \kappazero \Delta/2}/2 $.
%
It is worth noticing that in large surfaces, the self-coupling between different modes ($u\neq n$) is approximatively equal to zero, i.e., the orthogonality is preserved. Therefore, the only way to couple different modes is through the constitutive equations in \eqref{eq:JMs}.

\section{Constitutive Equations for Large Surfaces}
\label{Sec:D}

In this section, we show some derivation examples of matrix $\boldD$ in \eqref{eq:bmv}, associated with the constitutive equations in \eqref{eq:JMs}, for a large surface \ac{EMO}. 
Consider the generic $m$th \ac{EMO} with surface $\Scal^{(m)}$. In the following, we omit the superscript $m$ to lighten the notation.
To simplify the examples, we consider the isotropic case, then tensor $\Dw$ is of multiplicative type as  $\Dw =X_{\text{w}}(\boldr)$,
%
%
with $\text{w} \in \{ \text{JE}, \text{JH}, \text{ME}, \text{MH} \}$ and $\boldr \in \Scal$ \cite{AchSalCal:15}.  As for current sources and \ac{EM} fields,  function $X_{\text{w}}(\boldr)$ can be expressed in terms of series expansion 
\begin{equation} \label{eq:Xw}
X_{\text{w}}(\boldr)=\sum_{j=1}^{N_x \, N_y}  X_{\text{w}_{j}}  \, \phi_{j}(\boldr) \, e^{-\jmath \kappazero \, \Delta/2}
\end{equation}
where $X_{\text{w}_{j}}=\left < X_{\text{w}}(\boldr) , \phi_{j}(\boldr)\right >$ and $\left \{ \phi_j(\boldr) \right \}$ is a (scalar) basis set for $\Scal$.
The exponential term in \eqref{eq:Xw} accounts for the fact that we consider the currents located at $z=0$, whereas the \ac{EM} fields are observed on the right/left sides of the surface at $z=\pm \Delta/2$, according to the model described in Sec. \ref{Sec:surfacemodel}. 
%
%
It follows that matrix $\boldD_{\text{w}}$ in \eqref{eq:Dm}  can be written as $\boldD_{\text{w}}=\diag{\boldX_{\text{w}},\boldX_{\text{w}},\boldX_{\text{w}},\boldX_{\text{w}}}$, where $\boldX_{\text{w}}$ is a $N_x \times N_y$ matrix whose generic element is given by
\begin{align} \label{eq:xwun}
\left [  \boldX_{\text{w}} \right ]_{u,n}&=\innerprod{\Dw \cdot \Phim_u (\boldr)  }{\Phim_n (\boldr)} 
= \sum_{j=1}^{N_x\, N_y} X_{\text{w}_j} \innerprod{\phi_j(\boldr) \cdot \phi_u (\boldr)  }{\phi_n (\boldr)} 
\end{align}
for $n,u=1,2, \ldots, N_x \, N_y$. In a more compact form we can write
\begin{equation} \label{eq:Xw1}
\boldX_{\text{w}} = \sum_{j=1}^{N_x\, N_y} X_{\text{w}_j}  \, \boldH_j
\end{equation}
where $\left [\boldH_j \right ]_{n,u}= \innerprod{\phi_j(\boldr) \,  \phi_u(\boldr)}{ \phi_n (\boldr)}$.  In case the basis functions in Sec. \ref{Sec:surfacemodel} are used, it is  
\begin{align} \label{eq:Hjnu}
\left [\boldH_j \right ]_{n,u}&= \frac{1}{\sqrt{L_x\, L_y}} \delta_{j_x+u_x-n_x} \delta_{j_y+u_y-n_y}  e^{-\jmath \kappazero \, \Delta/2} \, .
\end{align}

It is evident from \eqref{eq:Hjnu} that the effect produced by a surface on induced currents as a function of the \ac{EM} field corresponds to a coupling between different modes  depending on the values of coefficients $X_{\text{w}_j}$ which characterize the behavior of the surface.

\subsection{Modeling the Equivalent Homogenized Boundary Conditions}


We here illustrate how the equivalent homogenized boundary conditions typically used to model metasurfaces can be accounted for in our framework.
Specifically, we model the surface as an inhomogeneous sheet of polarizable particles characterized by an electric surface impedance and magnetic surface admittance. 
This constitutes the homogenized model of the surface where the average electric and magnetic fields induce electric and magnetic currents generating a discontinuity of the \ac{EM} field between the two sides of the surface \cite{DiRDanTre:22,MarMac:22,AchSalCal:15}.
The corresponding boundary conditions are referred to as generalized sheet transition conditions. 
We consider the case in which the surface imposes an equivalent  homogenized  boundary condition of the type \cite{AsaAlbTcvRubRadTre:16,DiRDanTre:22}
\begin{align}  \label{eq:Boundary}
\left (
\begin{array}{c}
   \Jms(\boldr)   \\
    \Mms(\boldr)   \\  
     \end{array} 
\right )
&= \frac{1}{2}
\left (
\begin{array}{cc}
   \Yje(\boldr) &  \Yjh(\boldr)  \\
   \Yme(\boldr) & \Ymh(\boldr)  \\  
     \end{array} 
\right )
\cdot 
\left (
\begin{array}{c}
   \Emt^{+}(\boldr)+ \Emt^{-}(\boldr) \\
    \Hmt^{+} (\boldr)+ \Hmt^{-} (\boldr) \\  
     \end{array} 
\right )
= \frac{1}{2}
\left (
\begin{array}{cc}
   \Yje(\boldr) &  \Yjh(\boldr)  \\
   \Yme(\boldr) & \Ymh(\boldr)  \\  
     \end{array} 
\right )
\cdot 
\left (
\begin{array}{cccc}
   1 &   1 & 0 & 0  \\
   0 & 0 &  1 &  1   \\  
     \end{array} 
\right )
\left (
\begin{array}{c}
   \Emt^{+}(\boldr)\\
   \Emt^{-}(\boldr) \\
    \Hmt^{+} (\boldr)\\
     \Hmt^{-} (\boldr) \\  
     \end{array} 
\right )
\end{align}
where $\Emt^{+}(\boldr)$, $\Hmt^{+}(\boldr)$, $\Emt^{-}(\boldr)$, $\Hmt^{-}(\boldr)$ are the electric and magnetic tangential fields, respectively, at the right (+) and left (+) sides of the  surface, 
being $\Yje(\boldr)$ and $\Ymh(\boldr) $ the electric sheet admittance and magnetic sheet impedance, respectively. 
It follows that $X_{\text{w}}(\boldr) =\Ym_{\text{w}}(\boldr) $ with $\text{w} \in \{ \text{JE}, \text{JH}, \text{ME}, \text{MH} \}$. In terms of linear algebra formulation, the previous equivalent homogenized boundary conditions read
\begin{align} \label{eq:Dbound}
\boldb= \frac{1}{2} \left (
\begin{array}{cccc}
  \boldX_{\text{JE}}  & \boldzero_N & \boldX_{\text{JH}} & \boldzero_N \\
 \boldzero_N &  \boldX_{\text{JE}}  & \boldzero_N &  \boldX_{\text{JH}}  \\
   \boldX_{\text{ME}}  & \boldzero_N  & \boldX_{\text{MH}}  & \boldzero_N \\  
   \boldzero_N  &  \boldX_{\text{ME}}  &\boldzero_N &  \boldX_{\text{MH}}   \\  
     \end{array} 
\right ) \left (\begin{array}{cccc}
 \boldI_{2N}  &  \boldI_{2N}   & \boldzero_{2N} & \boldzero_{2N} \\
 \boldzero_{2N} & \boldzero_{2N} &     \boldI_{2N}   &  \boldI_{2N}   \\  
     \end{array} 
\right ) 
  \left ( \begin{array}{c}
   \bolde^{+} \\
    \bolde^{-} \\  
    \boldh^{+} \\
    \boldh^{-} \\  
     \end{array} 
\right )= \boldD \, \boldf
\end{align}
with $N= N_x \, N_y$, where $\bolde^{+}$ and $\bolde^{-}$ represent, respectively, the first and the second group of $2N$ elements of vector $\bolde$ associated with the two sides of the surface. The same meaning holds for $\boldh^+$and  $\boldh^-$.

\subsection{Impedance Sheet}

For a metasurface backed by a ground plane, the boundary conditions can be expressed in terms of an impenetrable equivalent impedance or admittance which relates the average tangential electric and magnetic fields on top of the surface \cite{ZhuWanDaiDebPoo:22}
\begin{align}
\Emt(\boldr) =\Zm(\boldr) \, \versorz \times \Hmt(\boldr)=\Zm(\boldr) \, \Jms (\boldr) \, .
\end{align} 
In this case, $\Jms (\boldr) =\Yje(\boldr)  \, \Emt(\boldr)$,  $\Mms(\boldr)=0$,   
 with $\Yje(\boldr)=\Zm(\boldr)^{-1}$. As a consequence, matrix $\boldD$ in \eqref{eq:Dbound} simplifies into
 \begin{equation}
 \boldD=\left (
\begin{array}{cccc}
  \boldX_{\text{JE}}  & \boldzero_N &  \boldzero_N & \boldzero_N \\
 \boldzero_N &  \boldX_{\text{JE}}  & \boldzero_N &   \boldzero_N \\
    \boldzero_N  & \boldzero_N  &  \boldzero_N  & \boldzero_N \\  
   \boldzero_N  &   \boldzero_N  &\boldzero_N &   \boldzero_N   \\  
     \end{array} 
\right ) \left (\begin{array}{cccc}
 \boldI_{2N}  &  \boldzero_{2N}   & \boldzero_{2N} & \boldzero_{2N} \\
 \boldzero_{2N} & \boldzero_{2N} &     \boldzero_{2N}  & \boldzero_{2N}   \\  
     \end{array} 
\right )  \, .
  \end{equation}

\section{EM Transfer Function Computation Examples}
\label{Sec:Examples}

We are now in the position of deriving the relationship between the linear algebra method illustrated in Sec. \ref{Sec:Linear} and the system \ac{EM} transfer function defined in Sec. \ref{Sec:STF}. 
Now, suppose one is interested in finding, for example,  the $xx$ component  $\Htfxx(\wavefreqx,\wavefreqy , 
\wavefreqxp,\wavefreqyp; z_{\text{s}}\, ,z_{\text{o}})$ of the \ac{EM} transfer function $\TildeGreenTot(\wavefreqx,\wavefreqy , 
\wavefreqxp,\wavefreqyp; z_{\text{s}}\, ,z_{\text{o}})$ in \eqref{eq:Htf}, which returns the system 
response observed on the plane at $z=z_{\text{o}}$  for the 2D wavenumber $(\wavefreqx,\wavefreqy)$ and polarization $\versora_x$ when solicited by the harmonic current in \eqref{eq:JPlanez}   located on the plane  $z=z_{\text{s}}$ with polarization $\versora=\versorx$. To this purpose, we add in the system two virtual \acp{EMO}  whose indexes are, respectively, $1$ and $M$, with $\boldD^{(1)}=\boldD^{(M)}=0$. The first \ac{EMO} is the responsible for the impinging  elementary harmonic electric current in \eqref{eq:JPlanez} with wavenumber $\wavefreqp$, whereas the $M$th \ac{EMO} is the virtual plane-wave \ac{EMO}, with wavenumber $\wavefreq$, given by \eqref{eq:Plane}. Therefore, the number of physical  \acp{EMO} is $M-2$.
The general problem is to find the algebrical relationship between $\fMv$ and $\aonev$ in \eqref{eq:fmv1}, where the coefficients in matrixes $\boldG^{(M,i)}$, for $i=2,3, \ldots, M-1$, are given by \eqref{eq:Gejkplane} and \eqref{eq:Gemkplane}, whereas the coefficients of $\boldG^{(m,1)}$, for $m=2,3,\ldots , M$, are given by \eqref{eq:Gejkplanecurr} and \eqref{eq:Gemkplanecurr}. 

Considering only the first component of $\eMv$ in $\fMv$ (that related to the $x$ polarization), it is 
\begin{align} \label{eq:Hxx}
\Htfxx(\wavefreqx,\wavefreqy , 
\wavefreqxp,\wavefreqyp; z_{\text{s}}\, ,z_{\text{o}})& =  \left [\eMv(\wavefreq,\wavefreqp) \right ]_{1,1}	
\end{align}
where vector $\eMv(\wavefreq,\wavefreqp)$ is derived by setting $\aonev=1$, 
and we made explicit the dependence of $\eMv$ on the wavenumbers $\wavefreqp$ and $\wavefreq$ of the impinging  elementary harmonic electric current and the observed wavenumber, respectively. A similar approach can be applied for the other polarizations.
In the following subsections, we illustrate some explicit examples of calculation of the \ac{EM} transfer function \eqref{eq:Hxx}.

\subsection{Transfer Function of a Finite Surface with Constant Impedance}

Suppose we want to find the transfer function of a generic \ac{EMO}, numbered with index 2, characterized by a given constitutive matrix $\boldD^{(2)}$. Here it is $M=3$. The observation and source planes are placed at $z_{\text{o}}=z_{\text{s}}=0$ and the surface at position $\boldp=(0,0,p_z)$ with $p_z>0$. By combining \eqref{eq:fmv1} for $m=2$ and $m=M=3$, it is
%
%
\begin{align} \label{eq:fM}
 	\fMv(\wavefreq,\wavefreqp)= &\boldf^{(3)}(\wavefreq,\wavefreqp)= \boldG^{(3,2)}(\wavefreq) \, \boldD^{(2)} \left ( \boldI -   \boldG^{(2,2)} \, \boldD^{(2)}  \right )^{-1}\,  \boldG^{(2,1)}(\wavefreqp) \, \aonev \, .
\end{align}

As an example, let the surface of the \ac{EMO} be characterized by a constant admittance across the surface $\Scal^{(2)}$ of size $\Lx \times \Ly$, i.e., $\Ym(\boldr)=Y=1/Z$, for $\boldr \in \Scal^{(2)}$, being $Z$ the surface's impedance.  From  \eqref{eq:xwun}, it follows that 
\begin{equation}
X_{\text{JE}_j} =\innerprod{\Ym(\boldr)}{\phi_j(\boldr)}=Y\, \sqrt{\Lx \, \Ly} \, \delta_{j-n_0}
\end{equation}
where $n_0=(N_x-1)/2+N_x (N_y-1)/2+1$ is the coefficient of the series expansion \eqref{eq:Xw} corresponding to the basis function for $n_x=n_y=0$ (continuous component). Therefore, only one term in \eqref{eq:Xw} is different from zero and matrix $\boldX_{\text{JE}}$ is given by $\boldX_{\text{JE}}=Y\, \sqrt{\Lx \, \Ly} \boldH_{n_0}$. By letting $\Delta \rightarrow 0$, it holds
\begin{align}
\boldD^{(2)}=Y\, 
\left (
\begin{array}{cccccccc}
  \boldI_N  & \boldzero_N &  \boldzero_N & \boldzero_N & \boldzero_N  & \boldzero_N &  \boldzero_N & \boldzero_N \\
 \boldzero_N &  \boldI_N & \boldzero_N &   \boldzero_N & \boldzero_N  & \boldzero_N &  \boldzero_N & \boldzero_N\\
    \boldzero_N  & \boldzero_N  &  \boldzero_N  & \boldzero_N & \boldzero_N  & \boldzero_N &  \boldzero_N & \boldzero_N\\  
   \boldzero_N  &   \boldzero_N  &\boldzero_N &   \boldzero_N   & \boldzero_N  & \boldzero_N &  \boldzero_N & \boldzero_N\\  
     \end{array} 
\right )   \, .
\end{align}
After a few tedious but straightforward matrix computations, it results 
\begin{align} \label{eq:htfxx}
\Htfxx & (\wavefreqx,\wavefreqy , 
\wavefreqxp,\wavefreqyp; 0,0)= 
\left [\eMv(\wavefreq,\wavefreqp) \right ]_{1}	
= \frac{ \eta^2   \, \Lx \, \Ly \, e^{\jmath p_z (k_z(\wavefreqx,\wavefreqy)+k_z(\wavefreqxp,\wavefreqyp))}}{ 4 }  \nonumber \\
 & \cdot \sum_{n_x=1}^{N_x} \sum_{n_y=1}^{N_y}  R_n\, \sinc{\frac{\wavefreqx \Lx}{2\pi}-n_x} \sinc{\frac{\wavefreqy \Ly}{2\pi}-n_y} \sinc{\frac{\wavefreqxp \Lx}{2\pi}-n_x} \sinc{\frac{\wavefreqyp \Ly}{2\pi}-n_y}
\end{align}
where $R_n=Y/ \left (1+ \frac{ Y\, \eta\, k_z^{(n)}}{2\, \kappazero\,    k_{n,n}^{(n)}} \right )$, $\eta=\kappazero/(\omega\, \epsilon)$, and we have exploited the following relationship $\versorx \scalprod  \left ( \crossprod{\wavefreq}{\crossprod{\wavefreq}{  \versorx }} \right )=k_z(\wavefreqx,\wavefreqy)$. From \eqref{eq:htfxx}, it can be evinced that each mode is, in general, subjected to a  different reflecting coefficient $R_n$. 

 It is interesting to investigate the particular case where $Z=0$, for which $R_n=2/\eta$.   Letting $N_x, N_y \rightarrow \infty$ and considering that $\sum_n \sinc{A-n} \, \sinc{B-n}=\sinc{A-B}$, \eqref{eq:htfxx} simplifies into
\begin{align} \label{eq:htfxx1}
\Htfxx & (\wavefreqx,\wavefreqy , 
\wavefreqxp,\wavefreqyp; 0,0)=
 \frac{ \eta   \, \Lx \, \Ly \, e^{\jmath p_z (k_z(\wavefreqx,\wavefreqy)+k_z(\wavefreqxp,\wavefreqyp))}}{ 2 }   
  \, \sinc{\frac{\Lx (\wavefreqx-\wavefreqxp) }{2\pi}} \sinc{\frac{\Ly (\wavefreqy -\wavefreqyp)}{2\pi}}  \, .
\end{align}
It is worth noticing that \eqref{eq:htfxx1} is proportional to the result found in \cite{NajJamSchPoo:21} related to the evaluation of the response of a finite-size rectangular perfect electric conductor.    
By letting $\Lx, \Ly \rightarrow +\infty$ and considering that $\lim_{x  \rightarrow +\infty} x \, \sinc{a\, x} = \delta(a)$, we obtain
\begin{align}  \label{eq:htfxx2}
\Htfxx & (\wavefreqx,\wavefreqy , 
\wavefreqxp,\wavefreqyp; 0,0)=
 \frac{ \eta \,  \,  e^{\jmath  2\, p_z k_z(\wavefreqx,\wavefreqy)}}{ 2 }  
  \cdot \delta(\wavefreqx-\wavefreqxp)  \,  \delta(\wavefreqy-\wavefreqyp) 
\end{align}
that is, the transfer function of an infinite size surface, recently derived in \cite{PizLozRanMar:23}, which represents a particular case of our more general formula  \eqref{eq:htfxx}. 
Specifically, \eqref{eq:htfxx2} indicates that the reflected field can be obtained equivalently by considering a virtual source at distance $2\, p_z$. This is nothing else than  the \emph{image theorem} saying that the reflection operated by a (large) perfect conductor is equivalent to a mirror image of the source   \cite{BalB:16}.

\subsection{RIS optimization}


In the next example, we consider a \ac{RIS} made of an inhomogeneous sheet of polarizable particles characterized by an electric surface impedance and magnetic surface admittance according to the equivalent  homogenized  boundary condition in \eqref{eq:Boundary}. 
Suppose that the \ac{RIS} has dimension $\Lx=1.06\,$m, $\Ly=1.06\,$m, and whose  
  purpose  is to reflect, with a reflection angle $\theta_\text{r}=22^{\circ}$ along the $x-z$ plane \big (i.e., $\wavefreqx^{(\text{r})}=\kappazero \sin(\theta_\text{r})\, , \, \wavefreqy^{(\text{r})}=0$\big ), an impinging \ac{EM} field with wavelength $\lambda=10\,$cm and incident angle $\theta_{\text{i}}=0$ \big ($\wavefreqxp^{(\text{i})}=\wavefreqyp^{(\text{i})}=0$\big ). Three different \ac{RIS} design methods are considered: \emph{Method 1)} Conventional approach where the surface is characterized by a periodic admittance $\Ym_{\text{JE}}(\boldr)=\frac{\jmath}{\eta} \sin \left ( \wavefreqx^{(\text{r})}  r_x \right) $,$\Ym_{\text{MH}}(\boldr)=\eta^2 \, \Ym_{\text{JE}}(\boldr)$, and $\Ym_{\text{ME}}(\boldr)=\Ym_{\text{JH}}(\boldr)=0$ \cite{AsaAlbTcvRubRadTre:16}.  Matrix $\boldD$ is designed as consequence using \eqref{eq:Xw}, \eqref{eq:Xw1}, \eqref{eq:Hjnu}, and \eqref{eq:Dbound};  
\emph{Method 2)} Matrix $\boldD$ is obtained as numerical solution of the following constrained nonlinear optimization problem:
\begin{align} \label{eq:simopt}
\max_{\boldD} & \left |\Htfxx  \left (\wavefreqx^{(\text{r})},\wavefreqy^{(\text{r})} , \wavefreqxp^{(\text{i})},\wavefreqyp^{(\text{i})}; 0,0 \right ) \right | 
& \text{s.t.} \, \, \, P_{\text{rad}}=\text{constant} 
\end{align}
with $|\Htfxx  \left (\wavefreqx^{(\text{r})},\wavefreqy^{(\text{r})} , \wavefreqxp^{(\text{i})},\wavefreqyp^{(\text{i})}; 0,0 \right )$ computed using  \eqref{eq:Hxx} and \eqref{eq:fM}; 
\emph{Method 3)} Matrix $\boldD$ is evaluated analytically by solving the following equation
$
 	 \boldD^{(2)} \left ( \boldI -   \boldG^{(2,2)} \, \boldD^{(2)}  \right )^{-1} = \boldT
$,
from which 
\begin{align} \label{eq:opt}
 	\boldD^{(2)}=  \left ( \boldI + \boldT\,  \boldG^{(2,2)}   \right )^{-1} \boldT
\end{align}
where $\boldT$ is the desired response of the \ac{RIS}. In particular, $\boldT$ is a zero matrix with only one element different from zero in the position where the input mode, corresponding to the impinging wave  $\wavefreqxp^{(\text{i})}=0$, is mapped to the output mode corresponding (or close) to  $\wavefreqx^{(\text{r})}$. Since we consider only the reflection along the $x-z$ plane,  the following numerical results were obtained by setting $N_x=25$ and $N_y=1$.

\begin{figure}[t]
\subfigure[]{\includegraphics[clip,width=0.49\columnwidth]{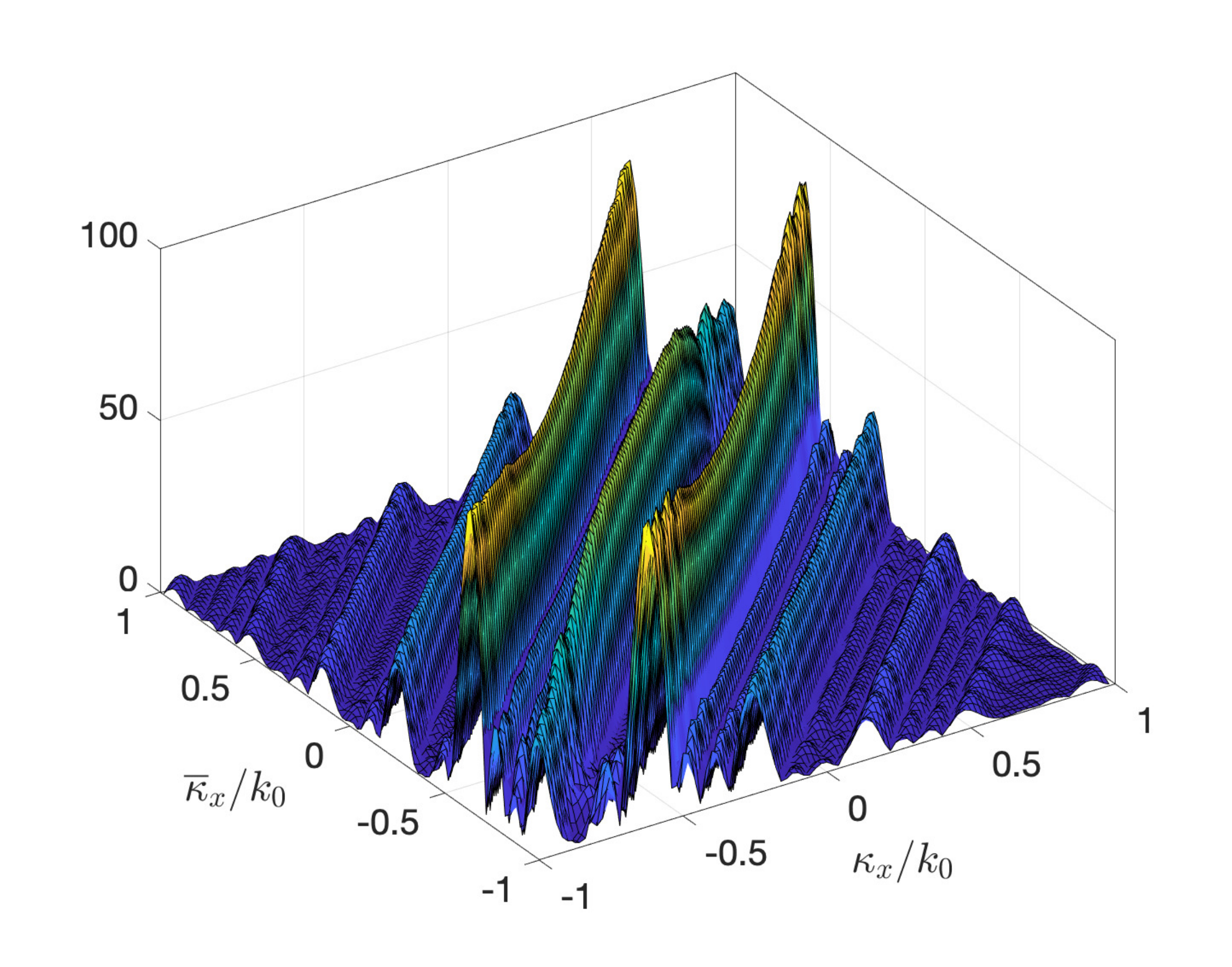}}
\subfigure[]{\includegraphics[clip,width=0.35\columnwidth]{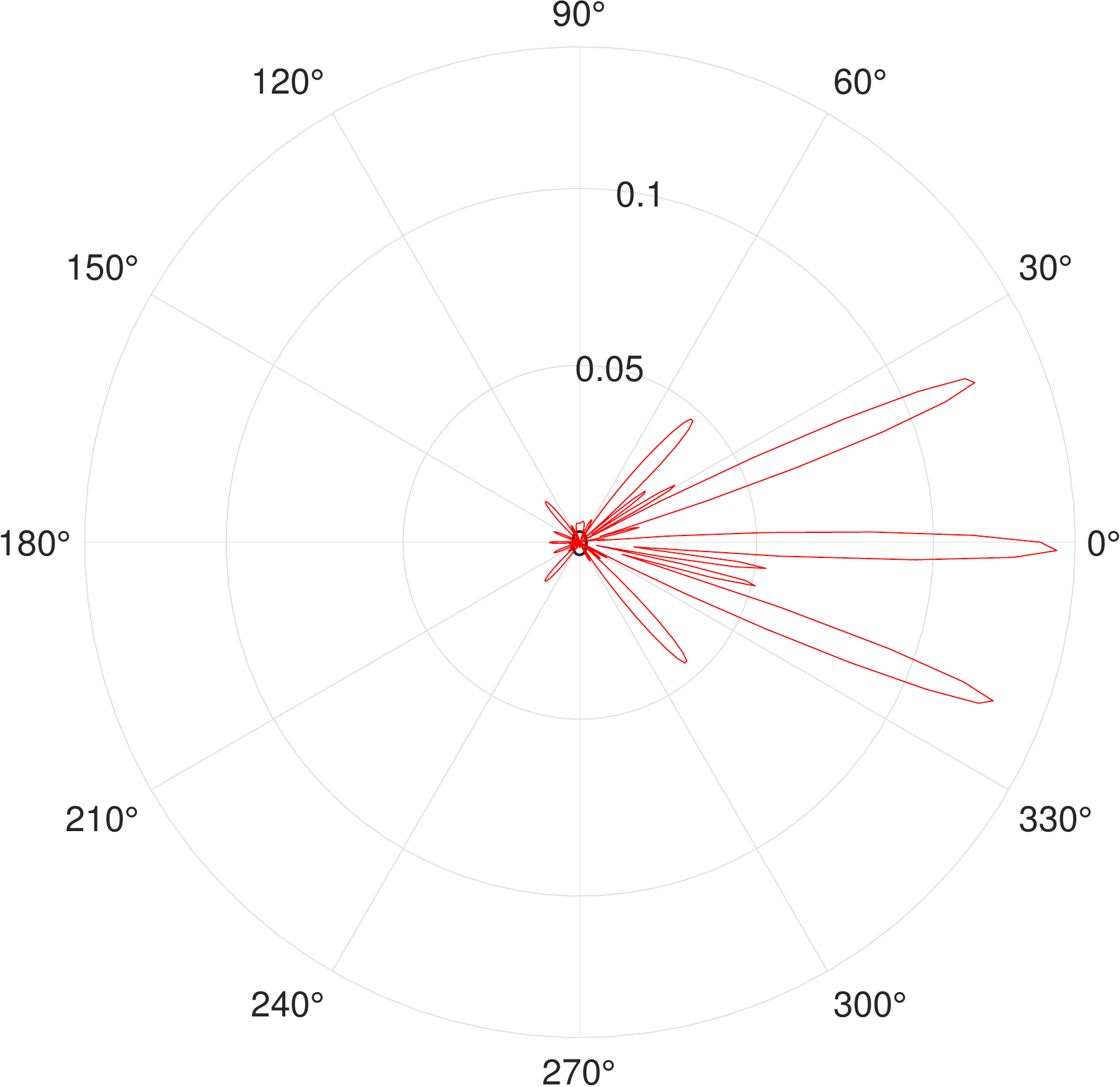}}
\caption{Method 1. (a) Amplitude of the \ac{RIS}' transfer function $\Htfxx  (\wavefreqx, 0 , \wavefreqxp; 0,0)$; (b) Radiation diagram of the \ac{RIS}. 
}
\label{Fig:Hsin}
\end{figure}

\begin{figure}[t]
\subfigure[]{\includegraphics[clip,width=0.49\columnwidth]{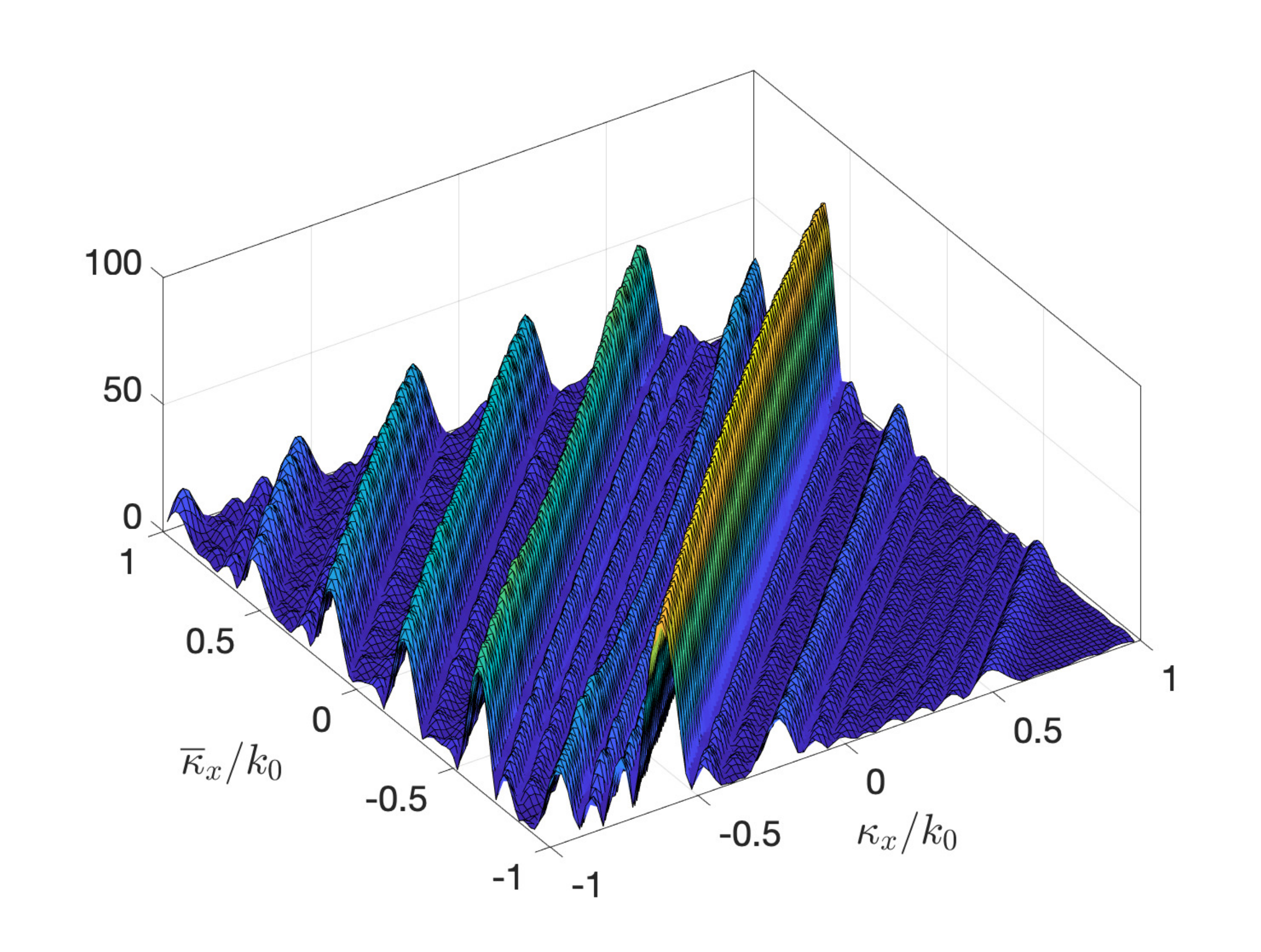}}
\subfigure[]{\includegraphics[clip,width=0.49\columnwidth]{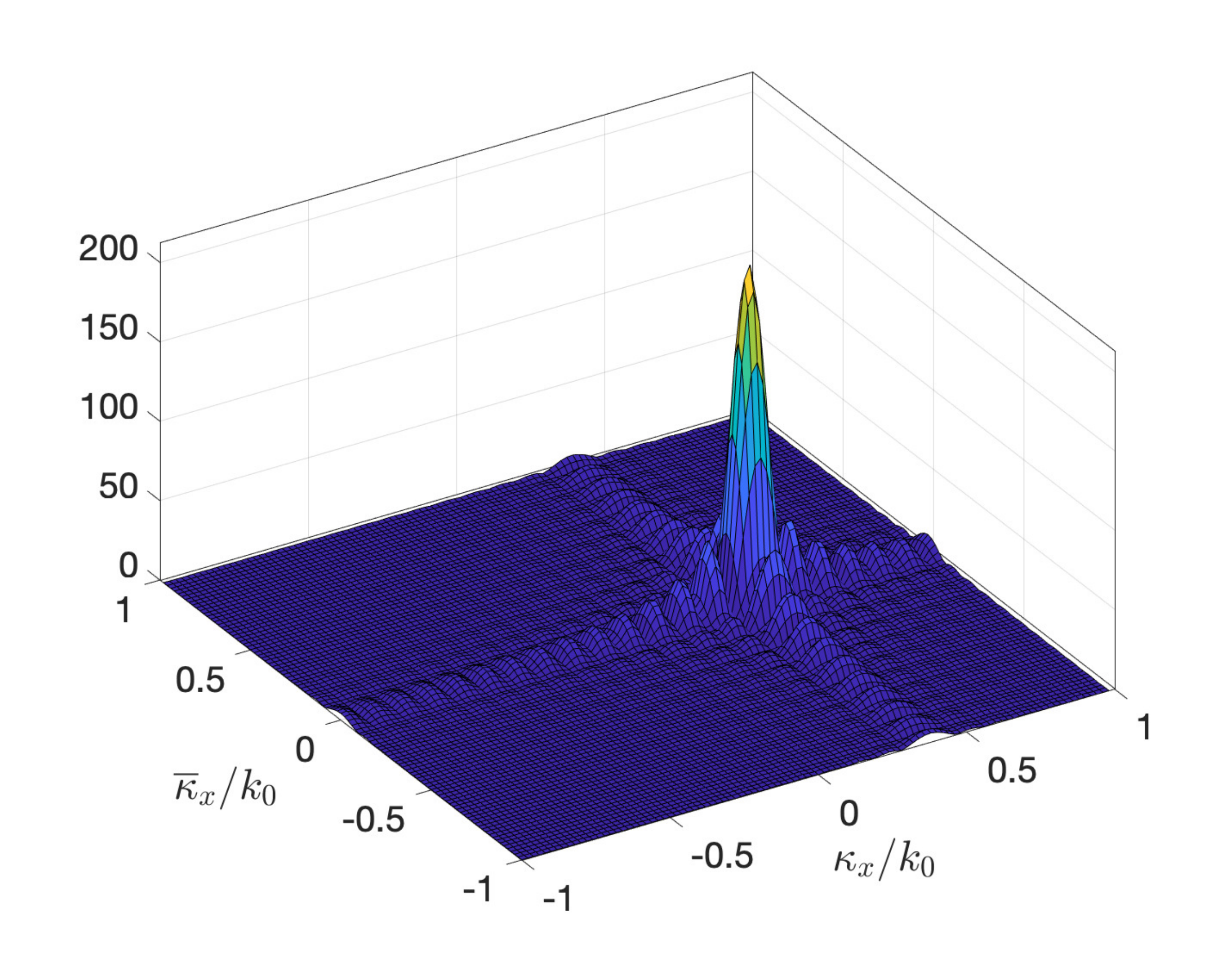}}
\caption{Amplitude of the \ac{RIS}' transfer function $\Htfxx  (\wavefreqx, 0 , \wavefreqxp; 0,0)$. (a) Method 2; (b) Method 3.}
\label{Fig:Hopt}
\end{figure}

In Fig.~\ref{Fig:Hsin}(a), the amplitude of the \ac{EM} transfer function $\Htfxx  (\wavefreqx, 0 , \wavefreqxp,; 0,0)$ for  a \ac{RIS} designed according to Method 1  is reported. The wavenumbers are normalized with respect to $\kappazero$. As expected, the transfer function provides some gain at $\wavefreqxp=0$ and $\wavefreqx=\wavefreqx^{(\text{r})}$ \big ($\wavefreqx^{(\text{r})}/\kappazero=0.38$\big ),  which means that the incident wave is correctly reflected towards $\theta_\text{r}$. This is evident also in Fig.~\ref{Fig:Hsin}(b), where the radiation diagram of the \ac{RIS} is reported. 
However, as it can be noticed in both figures, the periodic nature of the surface generates parasitic reflections in unwanted directions, as predicted by Floquet's theory whose evaluation  typically requires \ac{EM}-level simulations  \cite{DiaTre:21, DegVitDiRTre:22}.
When the \ac{RIS} is used in a  multi-user wireless system, such parasitic reflections  may generate interference to users located at different angles with respect to that of the target user. To reduce the interference, Method 2 can be adopted within our framework to strengthen the signal reflected in the right direction, thus reducing the intensity of the Floquet modes, as it can be noticed in Fig.~\eqref{Fig:Hopt}(a) obtained using Method 2. 
In any case, even if the Floquet modes are mitigated, the obtained \ac{EM} transfer function might still  generate significant interference. In fact, the off-diagonal behavior of the plot indicates that any other \ac{EM} wave arriving with a different incident angle, i.e., with $\wavefreqxp\neq 0$, would be reflected as well as that with $\wavefreqxp= 0$.
In other words, the \ac{RIS} acts as an anomalous mirror for all the signal sources present in the environment by generating additional interference in uncontrolled directions. This aspect has been often overlooked in the literature.  
Despite the optimization problem in \eqref{eq:simopt}  is aimed at eliminating any spurious reflection,  the particular structure of matrix $\boldD$, which is a linear combination of off-diagonal-like matrices $\boldH_j$ in \eqref{eq:Hjnu}, reduces the degree of freedom in designing $\boldD$ and prevents the achievement of the desired result. This constraint in the structure of matrix $\boldD$ is the consequence of the boundary conditions \eqref{eq:Boundary} that operate at a local level. 
Such a constraint is not  intrinsically present when using Method 3 according to which  $\boldD$ can take any form depending on the desired  behavior $\boldT$ from \eqref{eq:opt}. The corresponding \ac{EM} transfer function, depicted in Fig.~\ref{Fig:Hopt}(b), clearly reflects towards $\theta_{\text{r}}$ only those waves arriving with an incident angle $\theta_{\text{i}}=0$, whereas any other wave with different angle is not reflected, thus avoiding the generation of interference caused by Floquet modes and/or   other \ac{EM} sources.
From the practical point of view, an unconstrained matrix $\boldD$ requires more complex \ac{EM} structures imposing linear constraints at the global surface level \cite{TarEle:21}. Such structures, i.e., linear \acp{EMO}, act as \emph{universal mode converters}  \cite{Miller:12}.  %

\section{Conclusion}
\label{Sec:Conclusion}

In this paper, we have introduced a general and physically-consistent framework   for the characterization and design of programmable \ac{EM} environments under a signal processing perspective. Specifically, we have illustrated that any linear \ac{EM} environment in the presence of boundary conditions can be interpreted as a  space-variant linear feedback filter. We have then proposed a methodology  to characterize programmable \ac{EM} systems as a linear graph described by matrix operators, thus leading to the determination of the transfer function of the \ac{EM} system.  Finally, some examples of \ac{EM} transfer function computation and \ac{RIS} optimization methods have been provided to illustrate the utility of the proposed framework.

\section*{Acknowledgment}
This work was supported by the European Union under the Italian National Recovery and Resilience Plan (NRRP) of NextGenerationEU, partnership on ``Telecommunications of the Future" (PE00000001 - program ``RESTART"), and by the EU Horizon project TIMES (Grant no. 101096307).

\section*{Appendix}

Consider a source $\Am(\boldr)$, whose Fourier transform  $\TildeAm(\wavefreq)$ does not exhibit  singularities,  enclosed within the finite domain $\cal{D}$ such that $\Am(\boldr)=0$, $ \forall \boldr= (r_x,r_y,r_z) \notin {\cal{D}}$. Denote with $z_{\text{min}}=\min \left (r_z :   (r_x, r_y, r_z) \in {\cal{D}},  \forall  r_x, r_y\right )$ and with $z_{\text{max}}=\max \left (r_z :  (r_x,r_y,r_z) \in {\cal{D}}\, \forall  r_x, r_y \right )$. 
Suppose the purpose is to compute the inverse Fourier transform of $\TildeGreen(\wavefreq) \, \TildeAm(\wavefreq)$ at location $\boldr$, with $r_z>z_{\text{max}}$ or $r_z<z_{\text{min}}$. Unfortunately, $\TildeGreen(\wavefreq)$ in \eqref{eq:Greenk} presents a singularity when $|\wavefreq|=\kappazero$, i.e., the condition an \ac{EM} plane wave must satisfy. 
If we consider the integration over $\wavefreqz$ to be a contour integration with the contour completed at infinity, then the Cauchy's integral theorem has the effect of enforcing the condition $|\wavefreq|=\kappazero$ and hence the well-known more convenient representation of  the inverse Fourier transform can be found \cite{CleB:96}
\begin{align} \label{eq:Cauchy}
\frac{1}{(2 \pi)^3} \intthree \TildeGreen(\wavefreq)\, \TildeAm(\wavefreq) \, e^{\jmath \, \wavefreq \scalprod \boldr} \, d^3 \wavefreq=-\frac{\jmath \, \pi}{(2 \pi)^3}  \inttwo \frac{\TildeAm(\wavefreq^{\pm})}{k_z(\wavefreqx , \wavefreqy )}  \, e^{\jmath \, \wavefreq^{\pm} \scalprod \boldr}  \, d \wavefreqx \, d \wavefreqy
\end{align}
where 
\begin{equation} \label{eq:kz}
k_z=k_z(\wavefreqx,\wavefreqy)=\left \{ 
\begin{array}{cc}
\sqrt{\kappazero^2-\wavefreqx^2-\wavefreqy^2} & \, \, \, \, \, \left ( \wavefreqx, \wavefreqy \right ) \in \CalP \\
\jmath \sqrt{\wavefreqx^2+\wavefreqy^2-\kappazero^2} & \, \, \, \, \, \left ( \wavefreqx, \wavefreqy\right ) \notin \CalP
  \end{array} 
  \right .  
\end{equation}
\begin{equation} \label{eq:kpm}
\wavefreq^{\pm}=\left \{ 
\begin{array}{cc}
(\wavefreqx, \wavefreqy, k_z) & r_z > z_{\text{max}} \\
(\wavefreqx, \wavefreqy, -k_z) & r_z < z_{\text{min}}
  \end{array} 
  \right .  \, .
\end{equation}
and $\CalP=\left \{ \left ( \wavefreqx, \wavefreqy \right ) \in \mathbb{R}^2 : \wavefreqx^2+\wavefreqy^2 \le \kappazero^2 \right \}$. 
When $\left ( \wavefreqx, \wavefreqy\right ) \in \CalP$, $k_z$ is real and propagation happens. Instead, when  $\left ( \wavefreqx, \wavefreqy \right ) \notin \CalP$, $k_z$ is purely imaginary and the plane waves are evanescent.
Eqn. \eqref{eq:Cauchy} indicates that 3D Fourier-like integrals can be evaluated through  2D Fourier integrals, thus revealing that the \ac{EM} field representation in the 3D space has 2 degrees of freedom due to the Helmholtz equation the field has to satisfy accounted for by $\TildeGreen(\wavefreq)$ in \eqref{eq:Greenk}.  Moreover, it is worth noticing that for each condition in \eqref{eq:kpm}, the 2D integral includes only those  plane waves propagating into the half-space that does not contain the source, as well as the evanescent waves.


If $|\boldr| \gg \lambda$ and $\TildeAm(\wavefreq)$ in \eqref{eq:Cauchy} is slow varying around the value $\wavefreq_{\boldr}=\kappazero \, \versorr$,  through the method of stationary phase, the following approximation holds \cite{BalB:16}
\begin{align} \label{eq:Approx}
\inttwo \frac{\TildeAm(\wavefreq)}{k_z(\wavefreqx,\wavefreqy )}  \, e^{-\jmath \, \wavefreq \scalprod \boldr}  \, d \wavefreqx \, d \wavefreqy  \simeq \TildeAm(\wavefreq_{\boldr})  \inttwo  \frac{e^{-\jmath \, \wavefreq \scalprod \boldr}}{k_z(\wavefreqx,\wavefreqy )}    \, d \wavefreqx \, d \wavefreqy= \jmath 2 \pi \TildeAm(\wavefreq_{\boldr}) \frac{ e^{-\jmath \, \kappazero \, |\boldr|} }{ |\boldr|} 
\end{align}
where the last equality is known as Weyl's identity \cite{CleB:96}.

\ifCLASSOPTIONcaptionsoff
\fi

\bibliographystyle{IEEEtran}
\bibliography{IEEEabrv,BiblioDD,MetaSurfaces,EMInformationTheory,IntelligentSurfaces,MassiveMIMO,THzComm,Channels,EMTheory,WINS-Books,Vari}

\begin{thebibliography}{10}
\providecommand{\url}[1]{#1}
\csname url@samestyle\endcsname
\providecommand{\newblock}{\relax}
\providecommand{\bibinfo}[2]{#2}
\providecommand{\BIBentrySTDinterwordspacing}{\spaceskip=0pt\relax}
\providecommand{\BIBentryALTinterwordstretchfactor}{4}
\providecommand{\BIBentryALTinterwordspacing}{\spaceskip=\fontdimen2\font plus
\BIBentryALTinterwordstretchfactor\fontdimen3\font minus
  \fontdimen4\font\relax}
\providecommand{\BIBforeignlanguage}[2]{{%
\expandafter\ifx\csname l@#1\endcsname\relax
\typeout{** WARNING: IEEEtran.bst: No hyphenation pattern has been}%
\typeout{** loaded for the language `#1'. Using the pattern for}%
\typeout{** the default language instead.}%
\else
\language=\csname l@#1\endcsname
\fi
#2}}
\providecommand{\BIBdecl}{\relax}
\BIBdecl

\bibitem{DiRZapDebAloYueRosTre:20}
M.~Di~Renzo, A.~Zappone, M.~Debbah, M.-S. Alouini, C.~Yuen, J.~de~Rosny, and
  S.~Tretyakov, ``Smart radio environments empowered by reconfigurable
  intelligent surfaces: How it works, state of research, and the road ahead,''
  \emph{IEEE Journal on Selected Areas in Communications}, vol.~38, no.~11, pp.
  2450--2525, 2020.

\bibitem{BarHamLonMonRamVelAleBil:22}
M.~Barbuto, Z.~Hamzavi-Zarghani, M.~Longhi, A.~Monti, D.~Ramaccia, S.~Vellucci,
  A.~Toscano, and F.~Bilotti, ``Metasurfaces 3.0: A new paradigm for enabling
  smart electromagnetic environments,'' \emph{IEEE Transactions on Antennas and
  Propagation}, vol.~70, no.~10, pp. 8883--8897, 2022.

\bibitem{DarDec:J21}
D.~Dardari and N.~Decarli, ``Holographic communication using intelligent
  surfaces,'' \emph{IEEE Communications Magazine}, vol.~59, no.~6, pp. 35--41,
  June 2021.

\bibitem{BjoWymMatPopSanCar:22}
E.~Bjornson, H.~Wymeersch, B.~Matthiesen, P.~Popovski, L.~Sanguinetti, and
  E.~de~Carvalho, ``Reconfigurable intelligent surfaces: A signal processing
  perspective with wireless applications,'' \emph{IEEE Signal Processing
  Magazine}, vol.~39, no.~2, pp. 135--158, 2022.

\bibitem{AbrDarDiR:J21}
A.~Abrardo, D.~Dardari, and M.~Di~Renzo, ``Intelligent reflecting surfaces:
  Sum-rate optimization based on statistical position information,'' \emph{IEEE
  Transactions on Communications}, vol.~69, no.~10, pp. 7121--7136, Oct 2021.

\bibitem{JenWal:08}
M.~A. Jensen and J.~W. Wallace, ``Capacity of the continuous-space
  electromagnetic channel,'' \emph{IEEE Transactions on Antennas and
  Propagation}, vol.~56, no.~2, pp. 524--531, 2008.

\bibitem{Miller:19}
\BIBentryALTinterwordspacing
D.~A.~B. Miller, ``Waves, modes, communications, and optics: a tutorial,''
  \emph{Adv. Opt. Photon.}, vol.~11, no.~3, pp. 679--825, Sep 2019. [Online].
  Available: \url{http://aop.osa.org/abstract.cfm?URI=aop-11-3-679}
\BIBentrySTDinterwordspacing

\bibitem{ZhaShlGuiDarImaEld:J22}
H.~Zhang, N.~Shlezinger, F.~Guidi, D.~Dardari, M.~F. Imani, and Y.~C. Eldar,
  ``Beam focusing for near-field multiuser {MIMO} communications,'' \emph{IEEE
  Transactions on Wireless Communications}, vol.~21, no.~9, pp. 7476--7490,
  Sep. 2022.

\bibitem{DecDar:J21}
N.~Decarli and D.~Dardari, ``Communication modes with large intelligent
  surfaces in the near field,'' \emph{IEEE Access}, vol.~9, pp.
  165\,648--165\,666, 2021.

\bibitem{DiaTre:21}
A.~Díaz-Rubio and S.~A. Tretyakov, ``Macroscopic modeling of anomalously
  reflecting metasurfaces: Angular response and far-field scattering,''
  \emph{IEEE Transactions on Antennas and Propagation}, vol.~69, no.~10, pp.
  6560--6571, 2021.

\bibitem{OzdBjoLar:20}
{\"O}.~{{\"O}zdogan}, E.~{Bj{\"o}rnson}, and E.~G. {Larsson}, ``Intelligent
  reflecting surfaces: Physics, propagation, and pathloss modeling,''
  \emph{IEEE Wireless Communications Letters}, vol.~9, no.~5, pp. 581--585,
  2020.

\bibitem{Dar:J20}
D.~Dardari, ``Communicating with large intelligent surfaces: Fundamental limits
  and models,'' \emph{IEEE Journal on Selected Areas in Communications},
  vol.~38, no.~11, pp. 2526--2537, Nov 2020.

\bibitem{Tre:15}
\BIBentryALTinterwordspacing
S.~A. Tretyakov, ``Metasurfaces for general transformations of electromagnetic
  fields,'' \emph{Philosophical Transactions of the Royal Society A:
  Mathematical, Physical and Engineering Sciences}, vol. 373, no. 2049, p.
  20140362, 2015. [Online]. Available:
  \url{https://royalsocietypublishing.org/doi/abs/10.1098/rsta.2014.0362}
\BIBentrySTDinterwordspacing

\bibitem{AchSalCal:15}
K.~Achouri, M.~A. Salem, and C.~Caloz, ``General metasurface synthesis based on
  susceptibility tensors,'' \emph{IEEE Transactions on Antennas and
  Propagation}, vol.~63, no.~7, pp. 2977--2991, 2015.

\bibitem{DegVitDiRTre:22}
V.~Degli-Esposti, E.~M. Vitucci, M.~D. Renzo, and S.~A. Tretyakov,
  ``Reradiation and scattering from a reconfigurable intelligent surface: A
  general macroscopic model,'' \emph{IEEE Transactions on Antennas and
  Propagation}, vol.~70, no.~10, pp. 8691--8706, 2022.

\bibitem{AsaAlbTcvRubRadTre:16}
\BIBentryALTinterwordspacing
V.~S. Asadchy, M.~Albooyeh, S.~N. Tcvetkova, A.~D\'{\i}az-Rubio, Y.~Ra'di, and
  S.~A. Tretyakov, ``Perfect control of reflection and refraction using
  spatially dispersive metasurfaces,'' \emph{Phys. Rev. B}, vol.~94, p. 075142,
  Aug 2016. [Online]. Available:
  \url{https://link.aps.org/doi/10.1103/PhysRevB.94.075142}
\BIBentrySTDinterwordspacing

\bibitem{MarMac:22}
E.~Martini and S.~Maci, ``Theory, analysis, and design of metasurfaces for
  smart radio environments,'' \emph{Proceedings of the IEEE}, vol. 110, no.~9,
  pp. 1227--1243, 2022.

\bibitem{ZhuWanDaiDebPoo:22}
J.~{Zhu}, Z.~{Wan}, L.~{Dai}, M.~{Debbah}, and H.~V. {Poor}, ``{Electromagnetic
  Information Theory: Fundamentals, Modeling, Applications, and Open
  Problems},'' \emph{arXiv e-prints}, p. arXiv:2212.02882, Dec. 2022.

\bibitem{GraDiR:21}
G.~Gradoni and M.~Di~Renzo, ``End-to-end coupling aware communication model for
  reconfigurable intelligent surfaces: An electromagnetic-compliant approach
  based on mutual impedances,'' \emph{IEEE Wireless Communications Letters},
  pp. 1--1, 2021.

\bibitem{NajJamSchPoo:21}
M.~Najafi, V.~Jamali, R.~Schober, and H.~V. Poor, ``Physics-based modeling and
  scalable optimization of large intelligent reflecting surfaces,'' \emph{IEEE
  Transactions on Communications}, vol.~69, no.~4, pp. 2673--2691, 2021.

\bibitem{DiRDanTre:22}
M.~Di~Renzo, F.~H. Danufane, and S.~Tretyakov, ``Communication models for
  reconfigurable intelligent surfaces: From surface electromagnetics to
  wireless networks optimization,'' \emph{Proceedings of the IEEE}, vol. 110,
  no.~9, pp. 1164--1209, 2022.

\bibitem{PooBroTse:05}
A.~S.~Y. {Poon}, R.~W. {Brodersen}, and D.~N.~C. {Tse}, ``Degrees of freedom in
  multiple-antenna channels: a signal space approach,'' \emph{IEEE Transactions
  on Information Theory}, vol.~51, no.~2, pp. 523--536, Feb 2005.

\bibitem{PizSanMar:22}
A.~Pizzo, L.~Sanguinetti, and T.~L. Marzetta, ``Fourier plane-wave series
  expansion for holographic {MIMO} communications,'' \emph{IEEE Transactions on
  Wireless Communications}, vol.~21, no.~9, pp. 6890--6905, 2022.

\bibitem{PizSanMar:22a}
------, ``Spatial characterization of electromagnetic random channels,''
  \emph{IEEE Open Journal of the Communications Society}, vol.~3, pp. 847--866,
  2022.

\bibitem{PizLozRanMar:23}
A.~Pizzo, A.~Lozano, S.~Rangan, and T.~L. Marzetta, ``Wide-aperture {MIMO} via
  reflection off a smooth surface,'' \emph{IEEE Transactions on Wireless
  Communications}, pp. 1--1, 2023.

\bibitem{OliSalMas:22}
G.~{Oliveri}, M.~{Salucci}, and A.~{Massa}, ``{Generalized Analysis and Unified
  Design of EM Skins},'' \emph{arXiv e-prints}, p. arXiv:2207.08419, Jul. 2022.

\bibitem{MasBenDaRGouBaoOliPolRocSal:21}
\BIBentryALTinterwordspacing
A.~Massa, A.~Benoni, P.~Da~Ru', S.~K. Goudos, B.~Li, G.~Oliveri, A.~Polo,
  P.~Rocca, and M.~Salucci, ``Designing smart electromagnetic environments for
  next-generation wireless communications,'' \emph{Telecom}, vol.~2, no.~2, pp.
  213--221, 2021. [Online]. Available:
  \url{https://www.mdpi.com/2673-4001/2/2/14}
\BIBentrySTDinterwordspacing

\bibitem{BalB:16}
C.~A. Balanis, \emph{Antenna Theory: analysis and design}.\hskip 1em plus 0.5em
  minus 0.4em\relax New Jersey, USA: Wiley, 2016.

\bibitem{HarringtonBook:2001}
R.~F. Harrington, \emph{Time-Harmonic Electromagnetic Fields}.\hskip 1em plus
  0.5em minus 0.4em\relax New York, USA: IEEE Press - Wiley, 2001.

\bibitem{ChenB:18}
X.~Chen, \emph{Computational methods for electromagnetic inverse
  scattering}.\hskip 1em plus 0.5em minus 0.4em\relax Solaris South Tower,
  Singapore 138628: JohnWiley \& Sons Singapore Pte. Ltd,, 2018.

\bibitem{CleB:96}
P.~C. Clemmow, \emph{The Plane Wave Spectrum Representation of Electromagnetic
  Fields}.\hskip 1em plus 0.5em minus 0.4em\relax Walton Street, Oxford OX2
  6DPA: Oxford University Press, 1996.

\bibitem{Zad:50}
L.~Zadeh, ``Frequency analysis of variable networks,'' \emph{Proceedings of the
  IRE}, vol.~38, no.~3, pp. 291--299, 1950.

\bibitem{Har:67}
R.~Harrington, ``Matrix methods for field problems,'' \emph{Proceedings of the
  IEEE}, vol.~55, no.~2, pp. 136--149, 1967.

\bibitem{Rum:54}
\BIBentryALTinterwordspacing
V.~H. Rumsey, ``Reaction concept in electromagnetic theory,'' \emph{Phys.
  Rev.}, vol.~94, pp. 1483--1491, Jun 1954. [Online]. Available:
  \url{https://link.aps.org/doi/10.1103/PhysRev.94.1483}
\BIBentrySTDinterwordspacing

\bibitem{Mason:56}
S.~J. Mason, ``Feedback theory-further properties of signal flow graphs,''
  \emph{Proceedings of the IRE}, vol.~44, no.~7, pp. 920--926, 1956.

\bibitem{SheGloSanVar:89}
C.~Shen, K.~Glover, M.~Sancer, and A.~Varvatsis, ``The discrete {Fourier}
  transform method of solving differential-integral equations in scattering
  theory,'' \emph{IEEE Transactions on Antennas and Propagation}, vol.~37,
  no.~8, pp. 1032--1041, 1989.

\bibitem{TarEle:21}
\BIBentryALTinterwordspacing
S.~Taravati and G.~V. Eleftheriades, ``Programmable nonreciprocal meta-prism,''
  \emph{Scientific Reports}, vol.~11, no.~1, p. 7377, 2021. [Online].
  Available: \url{https://doi.org/10.1038/s41598-021-86597-1}
\BIBentrySTDinterwordspacing

\bibitem{Miller:12}
\BIBentryALTinterwordspacing
D.~A.~B. Miller, ``All linear optical devices are mode converters,'' \emph{Opt.
  Express}, vol.~20, no.~21, pp. 23\,985--23\,993, Oct 2012. [Online].
  Available:
  \url{http://www.osapublishing.org/oe/abstract.cfm?URI=oe-20-21-23985}
\BIBentrySTDinterwordspacing

\end{thebibliography}

\end{document}